\documentclass[floatfix,amsmath,amssymb,aps, prfluids]{revtex4-2}
\usepackage{amsmath}
\usepackage{bm}
\usepackage{appendix}
\usepackage{graphicx}
\usepackage{epstopdf}
\usepackage{float}
\usepackage{subcaption}
\usepackage{cleveref}
\usepackage{fullpage}
\usepackage{comment}
\usepackage{amsfonts}
\usepackage{mathrsfs}
\graphicspath{{figures/}}

\begin{document}

\title{Parameterization of hydrodynamic friction in a model for sheared suspensions of rough particles}

\author{Madhu V. Majji}
\email{madhuvr@mit.edu}
\author{James W. Swan}
\thanks{Deceased}
\affiliation{  Department of Chemical Engineering Massachusetts Institute of Technology, 
  \\ Cambridge, MA 02138, USA
}

\date{\today}

\begin{abstract}
In this work, we propose a method to parameterize a coarse grained, discrete element model for the hydrodynamic friction  between nearly touching rough spheres in suspension flows. The frictional resistance due to surface roughness primarily alters the sliding and rolling modes of motion of rough particles. Near contact Stokesian dynamics simulations incorporating a near-field pairwise resistance model accounting for these enhanced frictional modes was employed to compute particle trajectories in shear flow. In this model, the resistance to sliding and rolling modes of motion are augmented from a weakly diverging form proportional to log$(1/h)$, characteristic of smooth spheres, to a strongly diverging form scaling as $1/h$ for rough spheres, where $h$ is the mean surface to surface distance between nearly touching particles. The augmentation reflects the hydrodynamic resistance due to squeezing flows between surface asperities. We determine new bounds on the relative magnitude of the augmentations to the resistance to different modes of motion using inequality constraints reflecting the positive definiteness of the Stokes resistance tensor for a pair of the rough particles. Using the simulations of a particle pair in a shear flow, a simple model for angular rotation rate of the line of centers of the pair, $\omega$, is computed as a function of its orientation in the linear shear flow and the two free parameters of the augmented hydrodynamic resistance model:  the friction coupling strength, $\alpha$, and friction coupling range, $h_0$. Values of $\alpha$ and $h_0$ for rough particles synthesized experimentally can then be inferred by matching the pair rotation rate in the model to experimental observations of the same rotation rate when a dilute suspension of these rough particles is subjected to a linear shear flow. The same model is used to calculate the hydrodynamic contribution to the high frequency viscosity of rough particle suspensions. For different $\alpha$ and $h_0$ while assuming the particles otherwise behave as hard spheres, we observe that the viscosity diverges differently depending on the ratio of $h_0$ to the hydrodynamic radius of the particles.
\end{abstract}

\maketitle

\section{Introduction}
It is understood that fluid-particle suspensions encountered in nature and industry are often not well approximated with models for smooth spheres suspended in a Newtonian fluid when simulating their dynamics under flow. Wheat flour particles in bread making, casein micelles and whey protein particles in dairy industry, coffee particles in percolating coffee, corn starch particles in shear thickening suspensions, fumed silica or carbon black particles in colloidal gels and even highly engineered raspberry-like particles, to name a few, have surface asperities and jagged edges~\cite{xu2019high, mcmahon1998rethinking, alberghini2019coffee, galvez2017dramatic, barthel1999particle, kamp2016selective, lan2018unexpected, hsu2018roughness, hsiao2019experimental}. These rough features play a significant role in the rheology of these suspensions even when the scale of the asperities is small relative to the particle size.

Experiments reveal that sheared dense suspensions of rough particles display discontinuous shear thickening (DST) behaviour and rough particles suspensions jam under shear at lower particle concentration compared to the smooth particle suspensions. In such flows, when rough particles are nearly touching, interaction of asperities on the two surfaces leads to additional constraints on the sliding and rolling modes of the relative motion between the particles~\cite{fernandez2013microscopic, hsiao2017translational, hsu2018roughness, schroyen2019stress}. These near contact interactions are of at least two flavors when the particles are otherwise behaving as impenetrable: hydrodynamic interactions caused by the deformation of the fluid between the surface asperities of the two particles~\cite{jamali2019alternative, wang2020hydrodynamic}, and mechanical contact interactions as a result of asperities of the two particles physically touching each other~\cite{wyart2014discontinuous, seto2013discontinuous, mari2014shear, morris2018lubricated}. Theoretical and computational work in the past showed that each of the two interactions can independently produce the experimentally observed DST behavior in sheared suspensions of rough particles quite well~\cite{jamali2019alternative, wang2020hydrodynamic, seto2013discontinuous, mari2014shear}.

In the mechanical contact model, the resistance to sliding and rolling modes for a pair of touching rough particles arises from frictional forces based on the Coulomb's friction law in which the additional tangential forces due to the interaction of surface asperities is proportional to the normal loading. For identical spheres, this model breaks the time reversal symmetry of pair trajectories about the flow-gradient plane of the shear flow as the normal loading exists in the compression quadrant of the flow and is absent in the extensional quadrant~\cite{blanc2011experimental, gallier2014rheology}. This symmetry breaking occurs when the mean surface to surface separation between the particles is on the scale of the particle roughness which goes into the model as a length scale regulating the hydrodynamic lubrication.  

In the hydrodynamic model, the additional resistance to sliding and rolling modes of relative motion between nearly touching particles arises from an augmentation of the elements of the hydrodynamic resistance tensor corresponding to these modes. In particular one replaces the the weakly diverging $O($log$(1/h))$ contributions from sliding and rolling lubrication to a more strongly diverging $O(1/h)$ form, where $h$ is the mean surface to surface separation between the particles~\cite{wang2020hydrodynamic}. For spherical particles in this model, the symmetry of pair trajectories about the flow-gradient plane of the shear flow is preserved as the hydrodynamic frictional forces obey time reversal symmetry. Some other conservative force acting on the particles is needed to break the symmetry. 

To use these models with Stokesian dynamics simulations and simulate the flow of rough particle suspensions reflecting real world materials, one needs to be able to parameterize the model easily. Experimental measurements using atomic force microscopy (AFM), angle of repose method and cyclic shear cells, for example, were performed in the past to estimate the friction coefficient of the mechanical contact model for a wide range of particle dispersions as summarized in the recent review article by~\cite{hsiao2019experimental}. In atomic force microscopy measurements, to estimate the friction coefficient between rough particles, a particle attached to the AFM tip is dragged along a substrate made of the same material and the lateral force is measured while applying a fixed normal force. Note that this measurement gives a friction coefficient for non-rotating particles, whereas the particles rotate freely in sheared suspensions. The angle of repose method uses the steepest angle a pile of particles make with the horizontal floor to estimate the friction coefficient. This is a popular method used for granular flows and gives a measurement of static friction. Cyclic shear cells are used to measure, for an applied normal force, the bulk shear stress which can then be related to a bulk friction coefficient.

In this work, we focus on rough suspensions for which the hydrodynamic type of frictional interactions dominate the particle dynamics. A pair-wise hydrodynamic model based on the work of Wang~\textit{et al.}~\cite{wang2020hydrodynamic} was used to model the additional resistance due to interaction of asperities of rough particles in a suspension. The model was incorporated in to an in-house fast Stokesian dynamics software and simulated shear flow of suspensions of rough particles. The magnitude and the range of frictional resistance is controlled using two free parameters of the hydrodynamic friction model: the friction coupling strength $\alpha$ and the friction coupling range $h_0$. Wang~\textit{et al.}~\cite{wang2020hydrodynamic} in their simulation work used equal friction coupling strengths for a few selected resistance modes corresponding to sliding and rolling motions. We provide in this work, a frame work to establish the relative magnitudes of friction coupling strengths corresponding to various sliding and rolling modes of motion of a pair of rough particles. Using the simulations, a simple analytical model was developed to predict the rotation rate of a pair of rough particles interacting in shear flow as a function of the two free parameters of the hydrodynamic friction model. The parameters $\alpha$ and $h_0$ for real-world rough suspensions can be estimated by matching the experimentally observed pair-rotation rates of rough particles in shear flow with those predicted using the pair-rotation rate model.

\section{Simulation methodology}

\begin{figure}
 \centering
 \includegraphics[height=5cm]{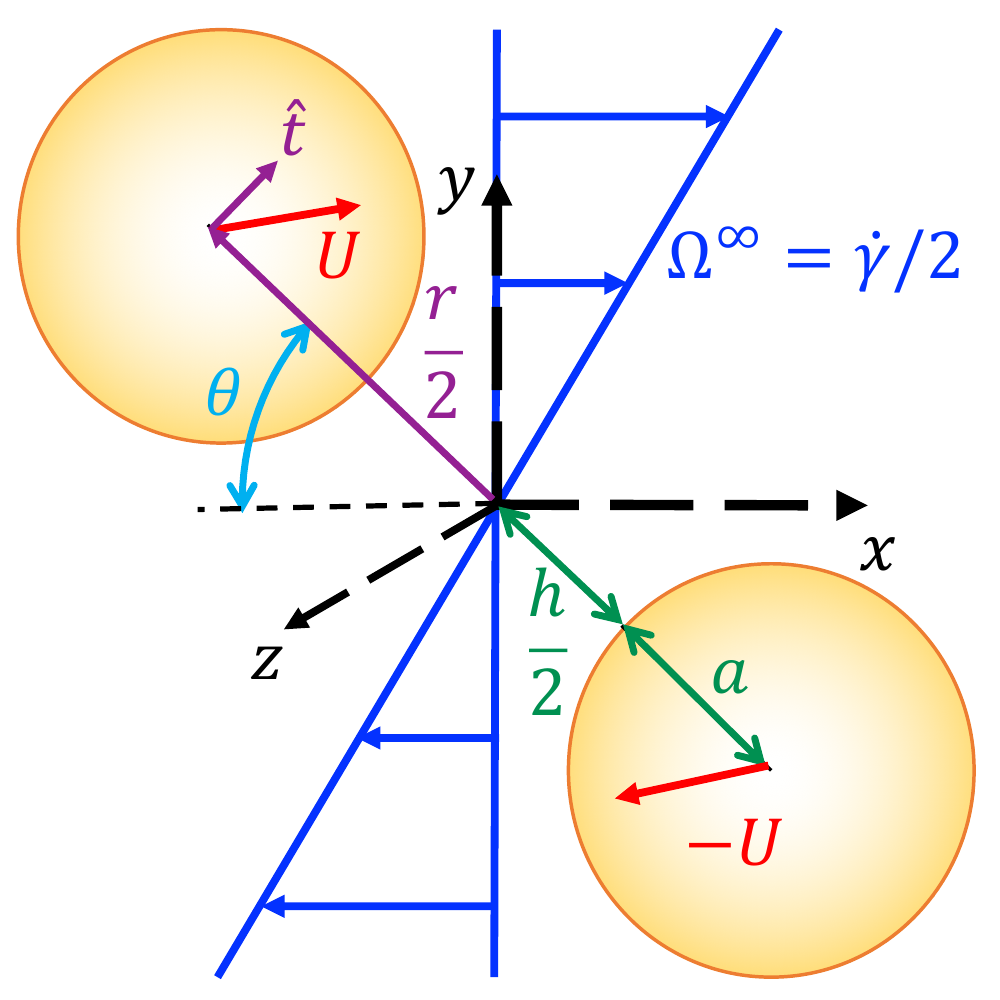}
 \caption{Schematic of a pair of particles in a simple shear flow.}
 \label{fgr:pairSchematic}
\end{figure}

The schematic in figure \ref{fgr:pairSchematic} shows a pair of rigid particles with spherical shape and hydrodynamic radii $a$ subjected to a simple shear flow.  The particles are assumed to be small and in a viscous liquid so that inertia is negligible. The flow has a vorticity of $\mathbf{\Omega^\infty}=(0,0,\dot{\gamma}/2)$ where $\dot{\gamma}$ is the shear rate.  The flow, gradient and vorticity directions of the shear field are aligned in $x$, $y$ and $z$ directions respectively. The positions of the two particles are $\mathbf{r_1}$ and $\mathbf{r_2}$.  The initial particle orientation is chosen to align with the flow-gradient plane such that the particle center line vector $\mathbf{r}=\mathbf{r_1}-\mathbf{r_2}$ makes an angle $\theta$ with the negative $x-$axis and has a magnitude $\lvert r \rvert = h+2a$ where $h$ is the separation between the hydrodynamic radii of the two particles. 

Following the standard formulation at low Reynolds number, the hydrodynamic forces $\mathbf{F}$, torques $\mathbf{L}$ and stresslets $\mathbf{S}$ on the particles are related linearly to the particle velocities $\mathbf{U}$ and angular velocities $\mathbf{\Omega}$ and applied strain rate $\mathbf{E}^{\infty}$ through the resistance tensor (equation \ref{resistaceeqns}). Here, $\mathbf{U}^{\infty}$ and  $\mathbf{\Omega}^{\infty}$ are  velocity and  vorticity of the applied flow. The components of the resistance tensor, $\mathbf{R}^{FU}$, for example, relate the hydrodynamic forces on different particles to the velocities of other particles. 

\begin{equation}
\begin{bmatrix}
\mathbf{F}\\
\mathbf{L}\\
\mathbf{S}
\end{bmatrix}
=-\begin{bmatrix}
\mathbf{R}^{FU} & \mathbf{R}^{F\Omega} & \mathbf{R}^{FE}\\
\mathbf{R}^{LU} & \mathbf{R}^{L\Omega} & \mathbf{R}^{LE}\\
\mathbf{R}^{SU} & \mathbf{R}^{S\Omega} & \mathbf{R}^{SE}
\end{bmatrix}
\begin{bmatrix}
\mathbf{U}-\mathbf{U}^{\infty}\\
\mathbf{\Omega}-\mathbf{\Omega}^{\infty}\\
-\mathbf{E}^{\infty}
\end{bmatrix}.
\label{resistaceeqns}
\end{equation}

The resistance matrix for a nearly touching particle pair can be constructed from the linear combination of various modes of relative particle motions, namely, particle translation and rotations parallel and transverse to their line of centers~\cite{jeffrey1984calculation,jeffrey1992calculation}. Since the extra frictional effects we seek to model act only on the modes of motion transverse to the particle center line and nonaxisymmetric modes of deformation, we consider explicitly the parts of the resistance tensors corresponding to these modes of motion and deformation. For the case of nearly touching vertically stacked particle pair orientation in linear shear flow ($\theta=90^{\circ}$ and $\mathbf{r}/\lvert r \rvert =(0,1,0)$), the resistance tensor takes the form in equation \ref{resistanceYs}. The equations are written for the two particles and only for the $x$ components of the forces ($F^1_x, F^2_x$), $z$ components of the torques ($L^1_x, L^2_x$) and $xy$ components of the stresslets ($S^1_{xy}, S^2_{xy}$) as the rest of the components are zero.  Here, $U^1_x-U_x^{\infty}(\mathbf{r_1})$, $U^2_x-U_x^{\infty}(\mathbf{r_2})$ are the $x$ components of the relative velocities, $\Omega_z^1-\Omega_z^{\infty}$, $\Omega_z^1-\Omega_z^{\infty}$ are the $z$ components of the relative angular velocities, $-E_{xy}^{\infty}$, $-E_{xy}^{\infty}$ are the $xy$ components of the relative strain rates of the two particles with respect to the background fluid, $Y^{AB}_{ij}$ is the corresponding scalar resistance function for the interaction of particle $i$ with particle $j$, $\mu$ is the viscosity of the fluid and $\mathscr{R}$ is the resistance tensor. Note that the scalar resistance coefficients $Y^A, Y^B, Y^G, Y^C, Y^H$ and $Y^M$ from the work of Jeffrey and Onishi~\cite{jeffrey1984calculation, jeffrey1992calculation} are replaced with $Y^{FU}, Y^{F\Omega}, Y^{FE}, Y^{L\Omega}, Y^{LE}$ and $Y^{SE}$, respectively, in this work.

\begin{equation}
\begin{gathered}
\begin{bmatrix}
F^1_x\\
F^2_x\\
L^1_z\\
L^2_z\\
S^1_{xy}\\
S^2_{xy}
\end{bmatrix}
=-\mathscr{R}
\begin{bmatrix}
U^1_x-U_x^{\infty}(\mathbf{r_1})\\
U^2_x-U_x^{\infty}(\mathbf{r_2})\\
\Omega_z^1-\Omega_z^{\infty}\\
\Omega_z^2-\Omega_z^{\infty}\\
-E_{xy}^{\infty}\\
-E_{xy}^{\infty}
\end{bmatrix},\\
\mathscr{R}=6\pi\mu a \begin{bmatrix}
Y^{FU}_{11} & Y^{FU}_{12} & \frac{2}{3}aY^{F\Omega}_{11} & -\frac{2}{3}aY^{F\Omega}_{12} & \frac{4}{3}aY^{FE}_{11} & -\frac{4}{3}aY^{FE}_{12}\\
Y^{FU}_{12} & Y^{FU}_{11} & \frac{2}{3}aY^{F\Omega}_{12} & -\frac{2}{3}aY^{F\Omega}_{11} & \frac{4}{3}aY^{FE}_{12} & -\frac{4}{3}aY^{FE}_{11}\\
\frac{2}{3}aY^{F\Omega}_{11} & \frac{2}{3}aY^{F\Omega}_{12} & \frac{4}{3}a^2Y^{L\Omega}_{11} & \frac{4}{3}a^2Y^{L\Omega}_{12} & -\frac{8}{3}a^2Y^{LE}_{11} &-\frac{8}{3}a^2Y^{LE}_{12}\\
-\frac{2}{3}aY^{F\Omega}_{12} & -\frac{2}{3}aY^{F\Omega}_{11} & \frac{4}{3}a^2Y^{L\Omega}_{12} & \frac{4}{3}a^2Y^{L\Omega}_{11} & -\frac{8}{3}a^2Y^{LE}_{12} & -\frac{8}{3}a^2Y^{LE}_{11}\\
\frac{2}{3}aY^{FE}_{11} & \frac{2}{3}aY^{FE}_{12} & -\frac{4}{3}a^2Y^{LE}_{11} & -\frac{4}{3}a^2Y^{LE}_{12} & \frac{10}{9}a^2Y^{SE}_{11} & \frac{10}{9}a^2Y^{SE}_{12}\\
-\frac{2}{3}aY^{FE}_{12} & -\frac{2}{3}aY^{FE}_{11} & -\frac{4}{3}a^2Y^{LE}_{12} & -\frac{4}{3}a^2Y^{LE}_{11} & \frac{10}{9}a^2Y^{SE}_{12} & \frac{10}{9}a^2Y^{SE}_{11}
\end{bmatrix}.
\end{gathered}
\label{resistanceYs}
\end{equation}

\subsection{Hydrodynamic friction model for rough particles}
To account for the effect of friction due to surface roughness on the pair particle dynamics, a hydrodynamic friction model based on the works of Jamali and Brady~\cite{jamali2019alternative} and Wang~\textit{et al.}~\cite{wang2020hydrodynamic}) is employed. The fundamental idea of this model is that when two rough surfaces slide past each other at close separation distances $h/a\ll1$, the fluid between surface asperities is subjected to compressive or squeezing deformations. In contrast, a pure sliding motion occurs between sliding smooth surfaces. Hence, the tangential hydrodynamics forces are modified from a weekly diverging form of $O($log$(1/h))$ for smooth surfaces to a strongly diverging form of $O(1/h')$ for roughened surfaces, where $h'$ is the distance between the asperities. This modification increases the resistance to sliding motion during both approach and departure, dissipating more energy during relative motion of nearly touching particles. This makes particles appear more tightly coupled in a shear flow.    

To model this increase in the sliding resistance due to asperities, an algebraic function $f_{ij}^{AB}$ that scales as $O(1/h)$ at close particle separations can be added to the scalar resistance functions $Y^{AB}_{ij}$ of the smooth particle cases (equation \ref{frictionFunction}). The two dimensionless free parameters of this model are $h_0/a$, which sets the length scale for onset of friction or the height of asperities relative to the particle radius, and $\alpha^{AB}_{ij}$, which sets the strength of friction or the density of asperities on the particle surface for the $AB$ resistance mode and for the interaction of particle $i$ with particle $j$. The function $f_{ij}^{AB}$, is chosen such that it smoothly approaches zero at $h = h_0$  and grows as $O(1/h)$ when $h \ll h_0$~\cite{wang2020hydrodynamic}. Note that when $\alpha^{AB}_{ij}=0$, the smooth particle case is recovered and for $\alpha^{AB}_{ij}\to\infty$ particle pairs become tightly locked together as though they are rigid dumbbells. Here, $\mathcal{H}( h_0 - h )$ is the Heaviside function which has a value of one for $h<h_0$ and zero otherwise. 

\begin{equation}
\begin{gathered}
    f_{ij}^{AB}(h)=\alpha^{AB}_{ij} \frac{a}{h_0}\left(\frac{h_0}{h}-3\frac{h}{h_0}+2\frac{h^2}{h_0^2} \right)\mathcal{H}(h_0-h),\\
    Y^{AB}_{ij}\rvert_{rough} = Y^{AB}_{ij}\rvert_{smooth} + f_{ij}^{AB}.\\
\end{gathered}
\label{frictionFunction}
\end{equation}

The special case of nearly touching rough particles can be used to constrain the relative magnitudes of different $\alpha^{AB}_{ij}$'s when $h/h_0 \ll 1 $.  In that condition, the contribution proportional to $ \alpha^{AB}_{ij}/h$ dominates the scalar resistance functions. Each of the scalar functions in equation \ref{resistanceYs} can be replaced with the corresponding $ \alpha^{AB}_{ij}/h$ multiplied by the correct power of the hydrodynamic radius $ a $. This approximate resistance tensor is given in equation \ref{resistanceYsfric}. 

\begin{equation}
\mathscr{R}=\frac{6\pi\mu a^2}{h}
\begin{bmatrix}
\alpha^{FU}_{11} & \alpha^{FU}_{12} & \frac{2}{3}a\alpha^{F\Omega}_{11} & -\frac{2}{3}a\alpha^{F\Omega}_{12} & \frac{4}{3}a\alpha^{FE}_{11} & \frac{4}{3}a\alpha^{FE}_{12}\\
\alpha^{FU}_{12} & \alpha^{FU}_{11} & \frac{2}{3}a\alpha^{F\Omega}_{12} & -\frac{2}{3}a\alpha^{F\Omega}_{11} & \frac{4}{3}a\alpha^{FE}_{12} & -\frac{4}{3}a\alpha^{FE}_{11}\\
\frac{2}{3}a\alpha^{F\Omega}_{11} & \frac{2}{3}a\alpha^{F\Omega}_{12} & \frac{4}{3}a^2\alpha^{L\Omega}_{11} & \frac{4}{3}a^2\alpha^{L\Omega}_{12} & -\frac{8}{3}a^2\alpha^{LE}_{11} &-\frac{8}{3}a^2\alpha^{LE}_{12}\\
-\frac{2}{3}a\alpha^{F\Omega}_{12} & -\frac{2}{3}a\alpha^{F\Omega}_{11} & \frac{4}{3}a^2\alpha^{L\Omega}_{12} & \frac{4}{3}a^2\alpha^{L\Omega}_{11} & -\frac{8}{3}a^2\alpha^{LE}_{12} & -\frac{8}{3}a^2\alpha^{LE}_{11}\\
\frac{2}{3}a\alpha^{FE}_{11} & \frac{2}{3}a\alpha^{FE}_{12} & -\frac{4}{3}a^2\alpha^{LE}_{11} & -\frac{4}{3}a^2\alpha^{LE}_{12} & \frac{10}{9}a^2\alpha^{SE}_{11} & \frac{10}{9}a^2\alpha^{SE}_{12}\\
-\frac{2}{3}a\alpha^{FE}_{12} & -\frac{2}{3}a\alpha^{FE}_{11} & -\frac{4}{3}a^2\alpha^{LE}_{12} & -\frac{4}{3}a^2\alpha^{LE}_{11} & \frac{10}{9}a^2\alpha^{SE}_{12} & \frac{10}{9}a^2\alpha^{SE}_{11}
\end{bmatrix}.
\label{resistanceYsfric}
\end{equation}

For an isolated pair particle system under linear shear flow, there must be symmetry about the mid-plane of the particle center line. This means that the $x$ components of the relative velocities of and forces on the two particles are equal in magnitude and opposite in direction, ${U^1_x-U_x^{\infty}(\mathbf{r_1}) = - (U^2_x-U_x^{\infty}(\mathbf{r_2}))}$ and ${F^1_x=-F^2_x}$. The $z$ components of the relative angular velocities of and torques on the two particles are equal, ${\Omega_z^1-\Omega_z^{\infty} = \Omega_z^2-\Omega_z^{\infty}}$ and ${L^1_z=L^2_z}$. The $xy$ components of the relative strain rates of and stresslets on the two particles are equal, ${-E_{xy}^{\infty}=-E_{xy}^{\infty}}$ and ${S^1_{xy}=S^2_{xy}}$. The symmetry condition on the force implies that the terms of the first row of the resistance tensor match the negative of the corresponding term in the second row, and we conclude that ${\alpha^{FU}_{11} = -\alpha^{FU}_{12}}$, ${\alpha^{F\Omega}_{11} = -\alpha^{F\Omega}_{12}}$ and ${\alpha^{FE}_{11} = -\alpha^{FE}_{12}}$. Similarly, the symmetry condition on the torque and stresslet reveal that ${\alpha^{L\Omega}_{11}=\alpha^{L\Omega}_{12}}$, ${\alpha^{LE}_{11}=\alpha^{LE}_{12}}$ and ${\alpha^{SE}_{11}=\alpha^{SE}_{12}}$. Using these symmetry relations and after subtracting the force equations and adding the torque equations and stresslet equations between the two particles,  a simplified linear set of equations as shown in equation \ref{resistanceYsfric1particle} emerges. 

\begin{equation}
\begin{bmatrix}
2F_x^1\\
2L_z^1\\
2S_{xy}^1
\end{bmatrix}
=-\frac{24\pi \mu a^2}{h}
\begin{bmatrix}
\alpha^{FU}_{11} & \frac{2}{3}a\alpha^{F\Omega}_{11} &  \frac{2}{3}a\alpha^{FE}_{11} \\
\frac{2}{3}a\alpha^{F\Omega}_{11}& \frac{4}{3}a^2\alpha^{L\Omega}_{11}& -\frac{4}{3}a^2\alpha^{LE}_{11}\\
\frac{2}{3}a\alpha^{FE}_{11} & -\frac{4}{3}a^2\alpha^{LE}_{11} & \frac{5}{9}a^2\alpha^{SE}_{11}
\end{bmatrix}
\begin{bmatrix}
U_x^1-U_x^{\infty}(\mathbf{r_1})\\
\Omega_z^1-\Omega_z^{\infty}\\
-2E_{xy}^{\infty}
\end{bmatrix}.
\label{resistanceYsfric1particle}
\end{equation}

From here on, the particle ID superscripts will be ignored and $\alpha^{AB}_{11}$'s will be replaced with $\alpha^{AB}$'s for simplicity. If the particles in the flow are otherwise force free and torque free, $F_x=0$ and $L_z=0$, the relative velocity, relative angular velocity of the particle with respect to the background fluid and the total stresslet for a rough particle pair can be obtained by solving  equation \ref{resistanceYsfric1particle}. 

\begin{equation}
    \begin{gathered}
        \frac{U_x-U_x^{\infty}(\mathbf{r})}{E_{xy}^{\infty}}=a\frac{4\alpha^{FE}\alpha^{L\Omega}+4\alpha^{F\Omega}\alpha^{LE}}{3\alpha^{FU}\alpha^{L\Omega}-\alpha^{F\Omega}\alpha^{F\Omega}},\\
        \frac{\Omega_z-\Omega_z^{\infty}}{E_{xy}^{\infty}}=\frac{2\alpha^{FE}\alpha^{F\Omega}+6\alpha^{LE}\alpha^{FU}}{\alpha^{F\Omega}\alpha^{F\Omega}-3\alpha^{FU}\alpha^{L\Omega}},\\
        \frac{S_{xy}}{E_{xy}^{\infty}}=\frac{8\pi\mu a^4}{h}\left(\frac{5}{3}\alpha^{SE}-\frac{4\alpha^{FE}\alpha^{FE}\alpha^{L\Omega}+12\alpha^{LE}\alpha^{LE}\alpha^{FU}+8\alpha^{FE}\alpha^{F\Omega}\alpha^{LE}}{3\alpha^{FU}\alpha^{L\Omega}-\alpha^{F\Omega}\alpha^{F\Omega}}\right).
    \end{gathered}
    \label{rigidpair}
\end{equation}

For large friction strengths and small particle separations ($\alpha^{AB}\to\infty$ or $ \alpha^{AB}  > 0 $ and $h/a\to0$), a pair of particles stacked on top of one another should follow the motion of a rigid dumbbell in simple shear flow and rotate with the vorticity of the flow~\cite{majumdar1972stokes}. The velocity and the angular velocity of the particle are $U_x=a\dot{\gamma}/2$ and $\Omega_z=\Omega^{\infty}$ respectively. The background flow at the particle location is $U_x^{\infty}(\mathbf{r})=a\dot{\gamma}$, the vorticity of the flow is $\Omega^{\infty}=\dot{\gamma}/2$ and the applied strain rate $E_{xy}^{\infty}=\dot{\gamma}/2$. Applying the rigid pair constraint along with the force free and torque free condition on the particles in equation \ref{resistanceYsfric1particle} yields the relations between friction strengths in this model which are independent of the interparticle separation.

\begin{equation}
\begin{gathered}
    \frac{\alpha^{FE}}{\alpha^{FU}}= -\frac{3}{4},\\
    \frac{\alpha^{LE}}{\alpha^{FU}}=\frac{1}{4}\frac{\alpha^{F\Omega}}{\alpha^{FU}},\\
    \frac{S_{xy}}{E^{\infty}}= \frac{8\pi \mu a^4}{h}\alpha^{FU}\left(\frac{3}{4}-\frac{5}{3}\frac{\alpha^{SE}}{\alpha^{FU}}\right).
\end{gathered}
\label{alpharelations}
\end{equation}

In addition to these relationships, the friction strengths are constrained by the physical requirement that particle motions should be purely dissipative.  Therefore, the resistance matrix should be positive semi-definite. It then follows that the determinant of the upper left $n \times n$ block of the matrix in equation \ref{resistanceYsfric1particle} should be non-negative for $n=1,2$ and $3$.  From the $n=1$ case, we learn that $\alpha^{FU}\ge0$. From the $n=2$ case, we learn that ${\alpha^{L\Omega}\alpha^{FU} \ge \left(\alpha^{F\Omega}\right)^2/3}$. Therefore $\alpha^{L\Omega}\ge0$ and if $\alpha^{L\Omega}=0$ then $\alpha^{F\Omega}=0$. There can be no force-rotation frictional coupling without a torque-rotation coupling. The positivity condition for $n=3$ is too complex to prove useful without simplification. 

To simplify, we choose the smallest $\alpha^{L\Omega}$ satisfying the $ n = 2 $ inequality [$\alpha^{L\Omega} = (\alpha^{F\Omega})^2/(3\alpha^{FU})$] and combine it with equations \ref{resistanceYsfric1particle} and \ref{alpharelations}, which yields a resistance matrix with two free parameters, $\alpha^{F\Omega}/\alpha^{FU}$ and $\alpha^{SE}/\alpha^{FU}$ as shown in equation \ref{resistance2}. The three eigenvalues of this resistance matrix are non-negative real numbers when the relation in equation \ref{semidefinite2} is satisfied.  These constraints limit the physical range of friction coefficients possible in this model. Any set of $\alpha'$s satisfying relations established above and giving a positive semi-definite resistance tensor should be physically admissible.  The rheology of the set of all such admissible models is uninvestigated so far, and we do not pursue that here.

\begin{equation}
\begin{bmatrix}
F_x\\
L_z\\
S_{xy}
\end{bmatrix}
=\frac{4\pi a}{h}\alpha^{FU}
\begin{bmatrix}
3 & 2a\frac{\alpha^{F\Omega}}{\alpha^{FU}} &  -\frac{3a}{2} \\
2a\frac{\alpha^{F\Omega}}{\alpha^{FU}}& \frac{4}{3}a^2\left(\frac{\alpha^{F\Omega}}{\alpha^{FU}}\right)^2& -a^2\frac{\alpha^{F\Omega}}{\alpha^{FU}}\\
-\frac{3a}{2} & -a^2\frac{\alpha^{F\Omega}}{\alpha^{FU}} & \frac{5}{3}a^2\frac{\alpha^{SE}}{\alpha^{FU}}
\end{bmatrix}
\begin{bmatrix}
U_x-U_x^{\infty}(\mathbf{r})\\
\Omega_z-\Omega_z^{\infty}\\
-2E_{xy}^{\infty}
\end{bmatrix}.
\label{resistance2}
\end{equation}

\begin{equation}
        \frac{\alpha^{SE}}{\alpha^{FU}} \ge a^2\frac{36a^2\alpha^{F\Omega}+81\alpha^{FU}}{80a^2\alpha^{F\Omega}+180\alpha^{FU}}.
    \label{semidefinite2}
\end{equation}

For the simulations discussed in this work, $\alpha^{FU}$ is replaced with $\alpha$, which is varied as a free parameter of the model. As there are not enough relations to establish relative magnitudes of all $\alpha^{AB}$'s, the following choices are made to reduce the dimensionality of the free parameter space. We are interested in developing a simple model to estimate the hydrodynamic friction between a pair of rough particles using experimentally measured translational velocities of the particles in shear flow. And, experimentally it is relatively easier to track the particle translational velocities compared to the particle rotation rates. Hence, we chose a model that ignores the rotational contribution to the hydrodynamic friction, $\alpha^{L\Omega}=0$ and $\alpha^{F\Omega}=0$.    
To satisfy equations \ref{alpharelations} and \ref{semidefinite2} we require that ${\alpha^{FE}=  -3\alpha^{FU}/4, \alpha^{LE}= 0}$ and ${\alpha^{SE}= 9\alpha^{FU}/20}$. Note that Wang~\textit{et al.}~\cite{wang2020hydrodynamic} in their hydrodynamic friction model considered the frictional contributions only from $FU, FE$ and $SE$ modes, just as in the present work. However they considered the case with equal friction strengths for all the considered modes, $\alpha^{FU}=\alpha^{FE}=\alpha^{SE}$ and in this work, we established the relative magnitudes of the various friction strengths using a set of physical constraints. Their model successfully predicted both the continuous shear thickening and the discontinuous shear thickening behaviours in dense suspensions under shear along with negative first and second normal stress differences.   Our model does the same even though the relationship among these friction strengths is slightly different.  

\subsection{Simulation protocol}
The hydrodynamic friction model described in the previous section was evaluated pairwise and added to the near-field resistance tensor in a Fast Stokasian dynamics simulation tool~\cite{fiore2019fast} and simulations were performed to compute the trajectories of a pair of particles subjected to a linear shear flow. The particles have no forces acting on them other than the hydrodynamic forces.  The imposed shear rate has a strength of $\dot{\gamma}$, which is used to make the time scale in the simulation dimensionless.  The inertia in the simulation is negligible. The particle pair is placed at the center of a cubic box with a side length of $40a$. Lees Edwards boundary conditions are applied in the gradient direction and periodic boundary conditions are applied in the flow and vorticity directions.  The particles are allowed to move until they pass one another in the shear flow, and their configurations and velocities are tracked along this path.  Because the particle velocity depends explicitly only on the configuration, every trajectory comprises a unique locus of points in space and no trajectories cross. 

For a given particle configuration ($h$ and $\theta$, refer to figure \ref{fgr:pairSchematic}) of a pair of particles with radii $a$ and a given set of friction parameters $\alpha$ and $h_0$, the simulation was used to compute the translational and rotational velocities of force free particles in simple shear. The velocities of the particles in the pair were used to compute the angular rotation rate of the pair center line in the vorticity direction ${\mathbf{\omega}(h,\theta) = 2\lvert\mathbf{U}.\hat{\mathbf{t}}\rvert/\lvert\mathbf{r}\rvert}$, where $\mathbf{U}$ is the particle velocity, $\mathbf{r}$ is the distance vector connecting particle centers and $\hat{\mathbf{t}}$ is the unit vector normal to the particle center line. The pair angular velocity is sensitive to $\alpha$ as the increase in sliding resistance locks the particles together more strongly, which results in a more rigid-body-like motion of the pair. The simulations were performed for a range of values of friction model parameters: friction coupling strength between $\alpha=0$ and $\alpha=10^6$ and friction coupling range between $h_0/a=0.001$ and $h_0/a=1.0$. For each set of values of friction parameters, the pair rotation rate was computed for multiple particle separations between $h/a=10^{-4}$ and $h/a=h_0/a$ and for various pair orientations in the compression quadrant of the shear flow between $\theta = 0^{\circ}$ and $\theta = 90^{\circ}$. The simulation data was used to construct a simple model for the angular velocity of a rough particle pair, $\omega$, which depends on the configuration $(h/a,\theta)$ and the frictional parameters $(\alpha,h_0/a)$. This model along with particle trajectories recorded from a simple shear flow experiment of dilute suspension in a microfluidic device, for example, can be used to characterize the effect of asperities between a given rough particle pair in flow.

\section{Results and discussion}

\subsection{Dynamics of pairs aligned in the gradient direction}

\begin{figure}
     \centering
     \begin{subfigure}[b]{0.4\textwidth}
         \centering
         \includegraphics[width=\textwidth]{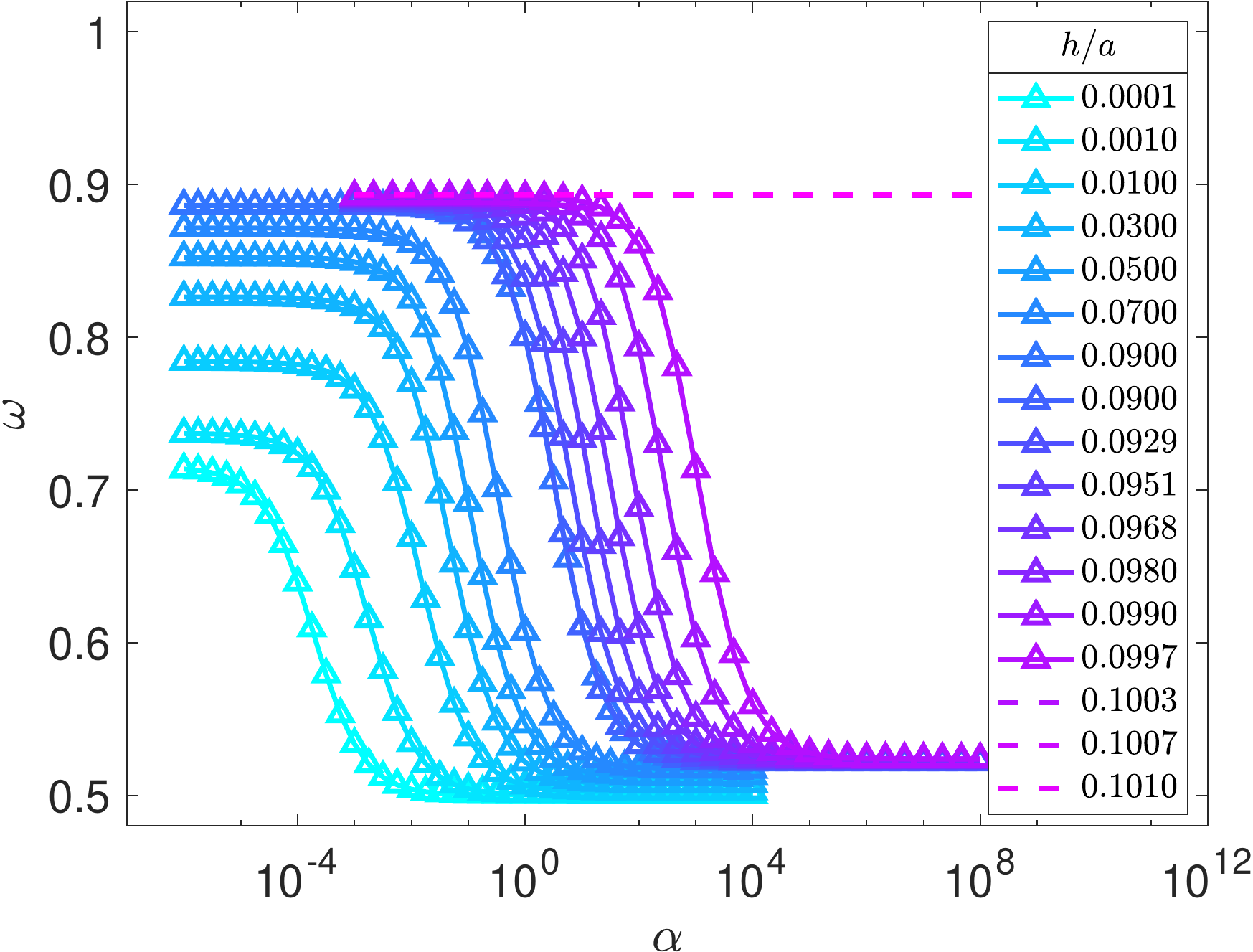}
         \caption{ }
         \label{fgr:OmegaAlpha90}
     \end{subfigure}
     \begin{subfigure}[b]{0.4\textwidth}
         \centering
         \includegraphics[width=\textwidth]{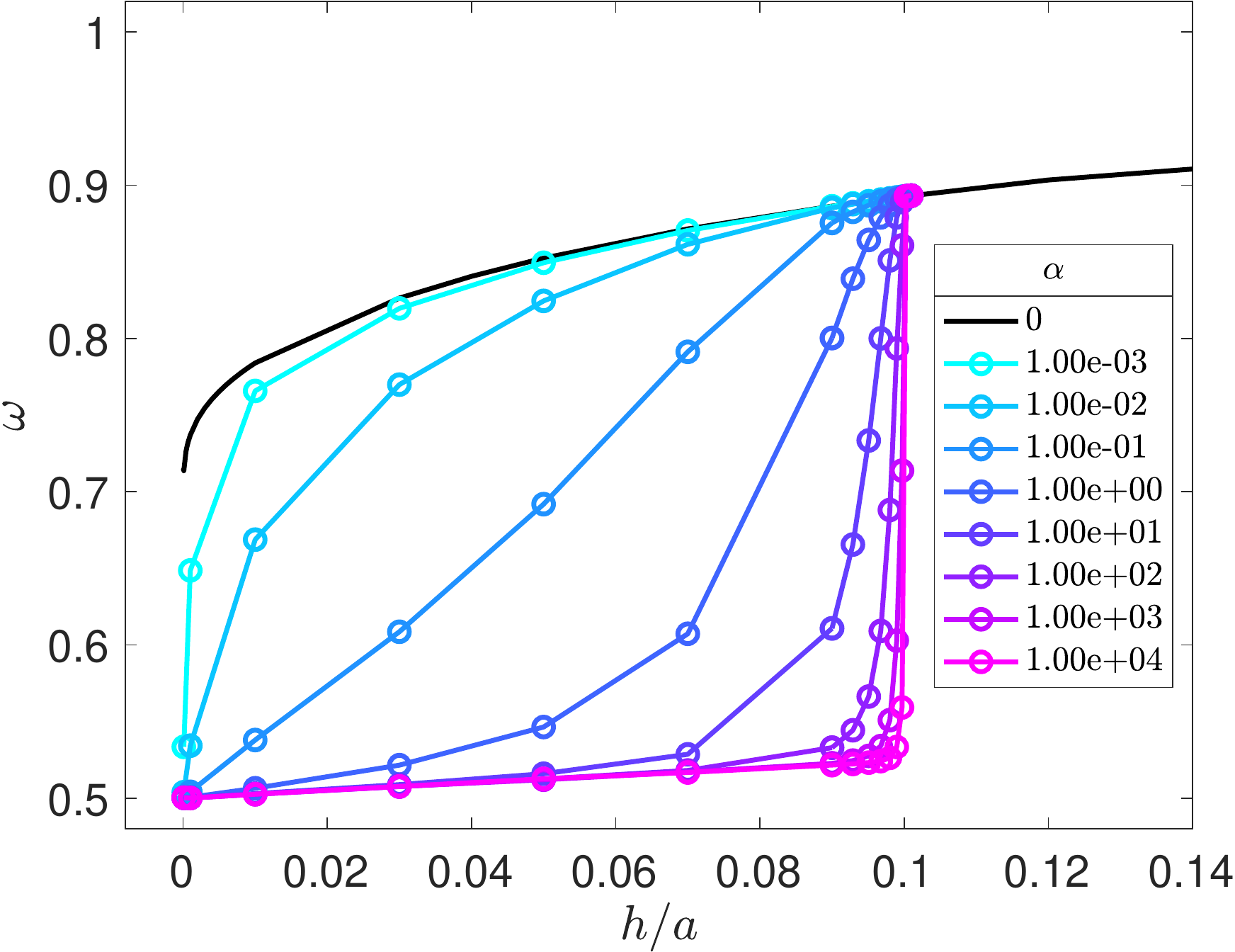}
         \caption{ }
         \label{fgr:Omegah90}
     \end{subfigure}
        \caption{Pair rotation rate $\omega$ as a function of (a) the friction strength $\alpha$ and (b) the separation distance $h/a$ between the particle surfaces for a friction range of $h_0/a=0.1$ when the particle centers are aligned in the gradient direction ($\theta=90^{\circ}$). The solid black line in (b) corresponds to smooth particle limit.}
        \label{fgr:theta90h00p1}
\end{figure}

Figures \ref{fgr:OmegaAlpha90} and \ref{fgr:Omegah90} show the rotation rate of the particle pair, $\omega$, when the particle centers are oriented along the gradient direction, $\theta=90^{\circ}$, in the shear flow and for a friction activation range of $h_0/a=0.1$. When the separation distance between the particle surfaces is within the friction range, $0<h/a<h_0/a$, the pair rotation rate as a function of the friction strength $\alpha$ follows an inverted sigmoid behavior with distinct $ h/a $ dependent asymptotes when $ \alpha $ is large and small as shown in the figure \ref{fgr:OmegaAlpha90}. As $\alpha$ approaches zero, pair rotation rate corresponding to smooth particles is recovered. The pair rotation rate is significantly larger than the rate of rotation of the imposed shear flow ($\Omega_{shear}=\dot{\gamma}/2$) as the particles can slide past each other easily. Whereas, at large values of $\alpha$ the pair rotates with a rate that is close to, and only slightly faster than, the rotation rate of the shear flow. The large sliding resistance arising from large $\alpha$ locks the particles into a rigid body. On decreasing $\alpha$ to an intermediate value, the pair rotation rate grows. For particle separations beyond the friction range, $h/a>h_0/a$, the smooth particle behaviour is recovered as expected and as shown by the dotted lines in the figure \ref{fgr:OmegaAlpha90}. The dependence of pair rotation rate $\omega$ on the particle separation $h/a$ is shown in figure \ref{fgr:Omegah90}. In the smooth particle limit, $\alpha\to0$, as the particle separation is reduced, the pair rotation rate reduces rather slowly and never reaches the rotation rate of the fluid. On the other hand, in the large $\alpha$ limit, the pair rotation rate nearly matches the rotation rate of the background flow over the range $0<h/a<h_0/a$. At $h/a\approx h_0/a$, the rotation rate jumps up sharply to the rotation rate corresponding to the smooth particle case and stays on the smooth particle curve for $h/a>h_0/a$.  This jump occurs where the hypothesized surface asperities no longer overlap. The curves for the intermediate $\alpha$ can be divided into two groups: concave up curves which result in friction dominated particle interactions, and concave down curves where the smooth particle lubrication interactions dominate.  

\begin{figure}
     \centering
     \begin{subfigure}[b]{0.3\textwidth}
         \centering
         \includegraphics[width=\textwidth]{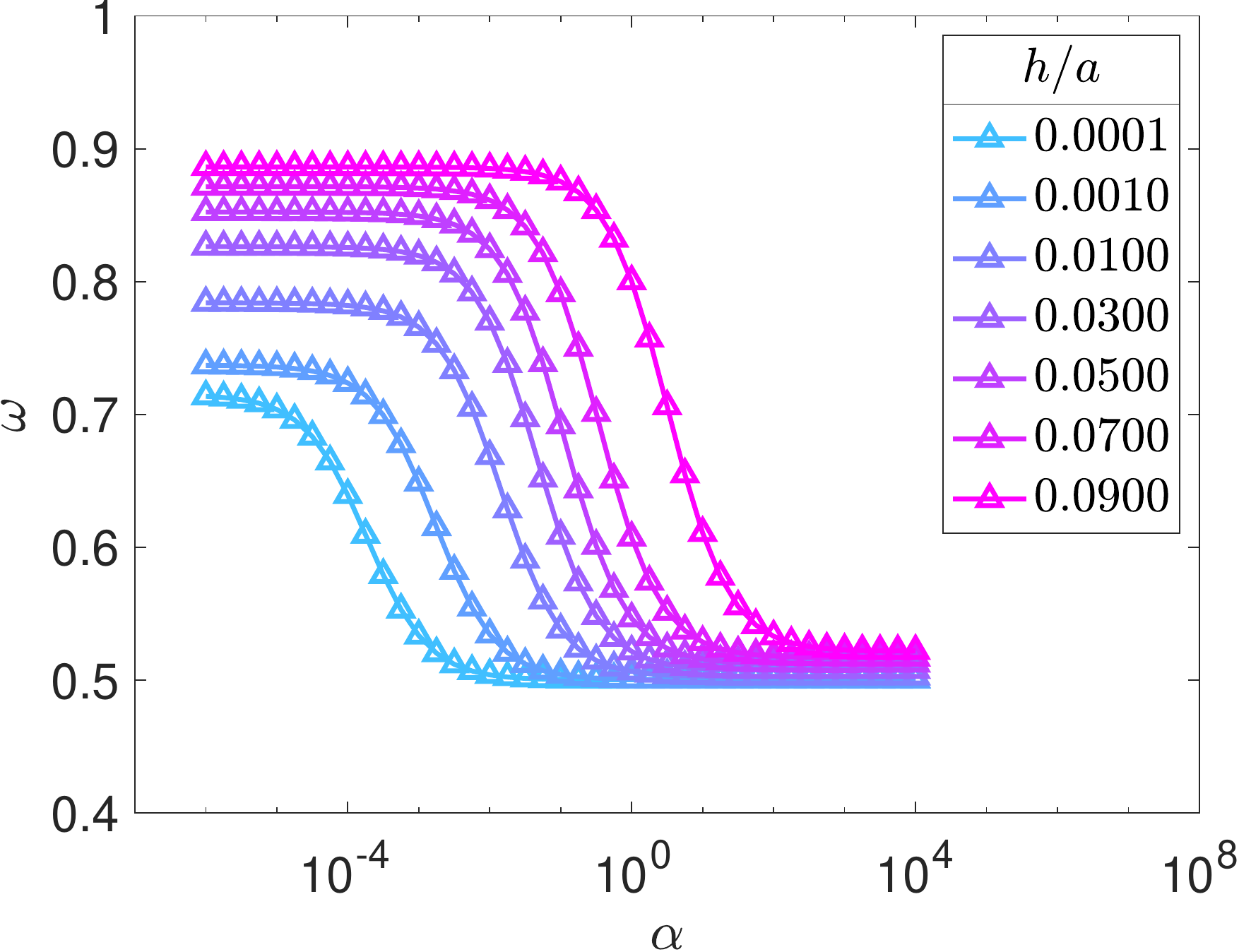}
         \caption{$h_0/a=0.1$}
         \label{fgr:OmegaAlpha90h00p1}
     \end{subfigure}
     \begin{subfigure}[b]{0.3\textwidth}
         \centering
         \includegraphics[width=\textwidth]{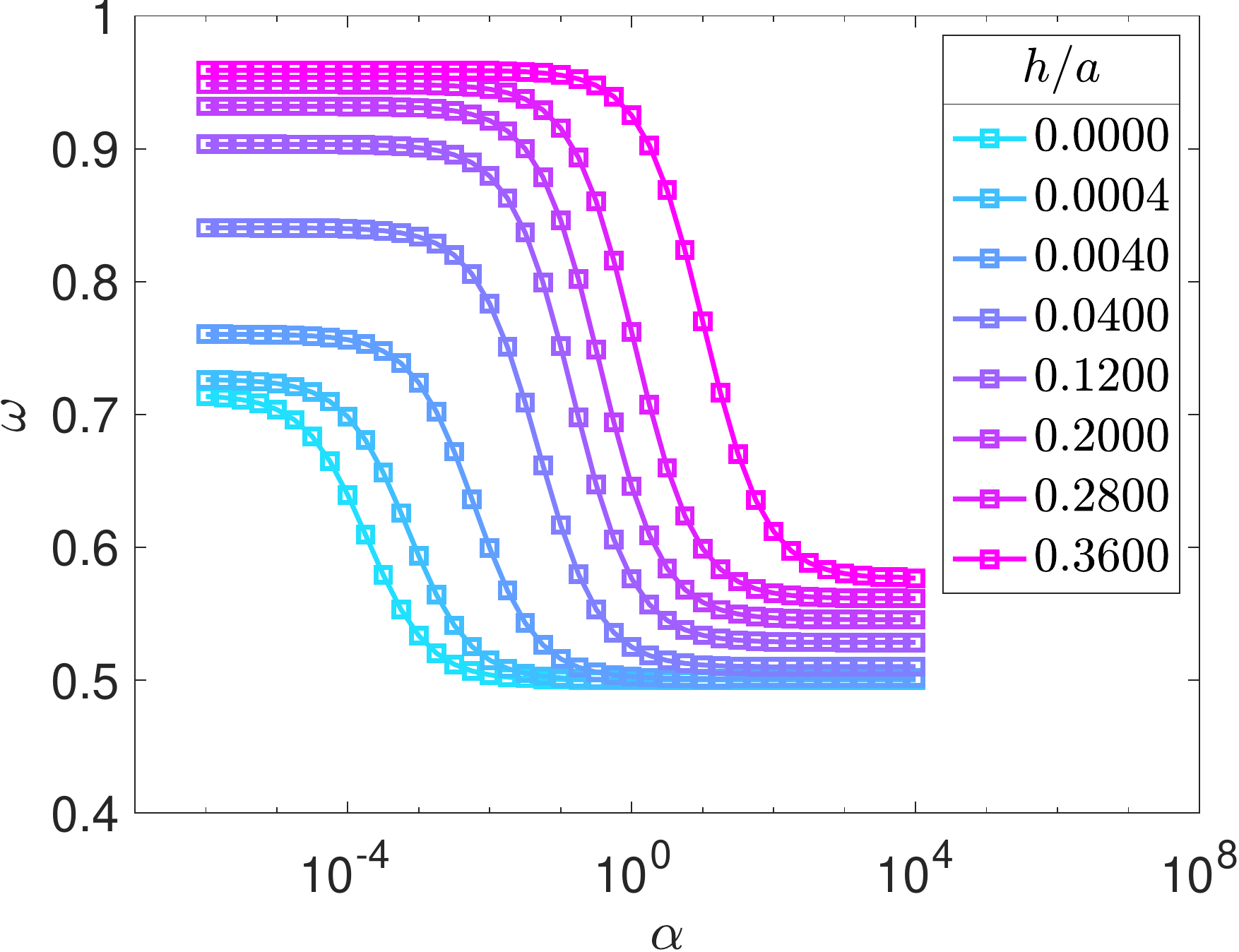}
         \caption{$h_0/a=0.4$}
         \label{fgr:OmegaAlpha90h00p4}
     \end{subfigure}
     \begin{subfigure}[b]{0.3\textwidth}
         \centering
         \includegraphics[width=\textwidth]{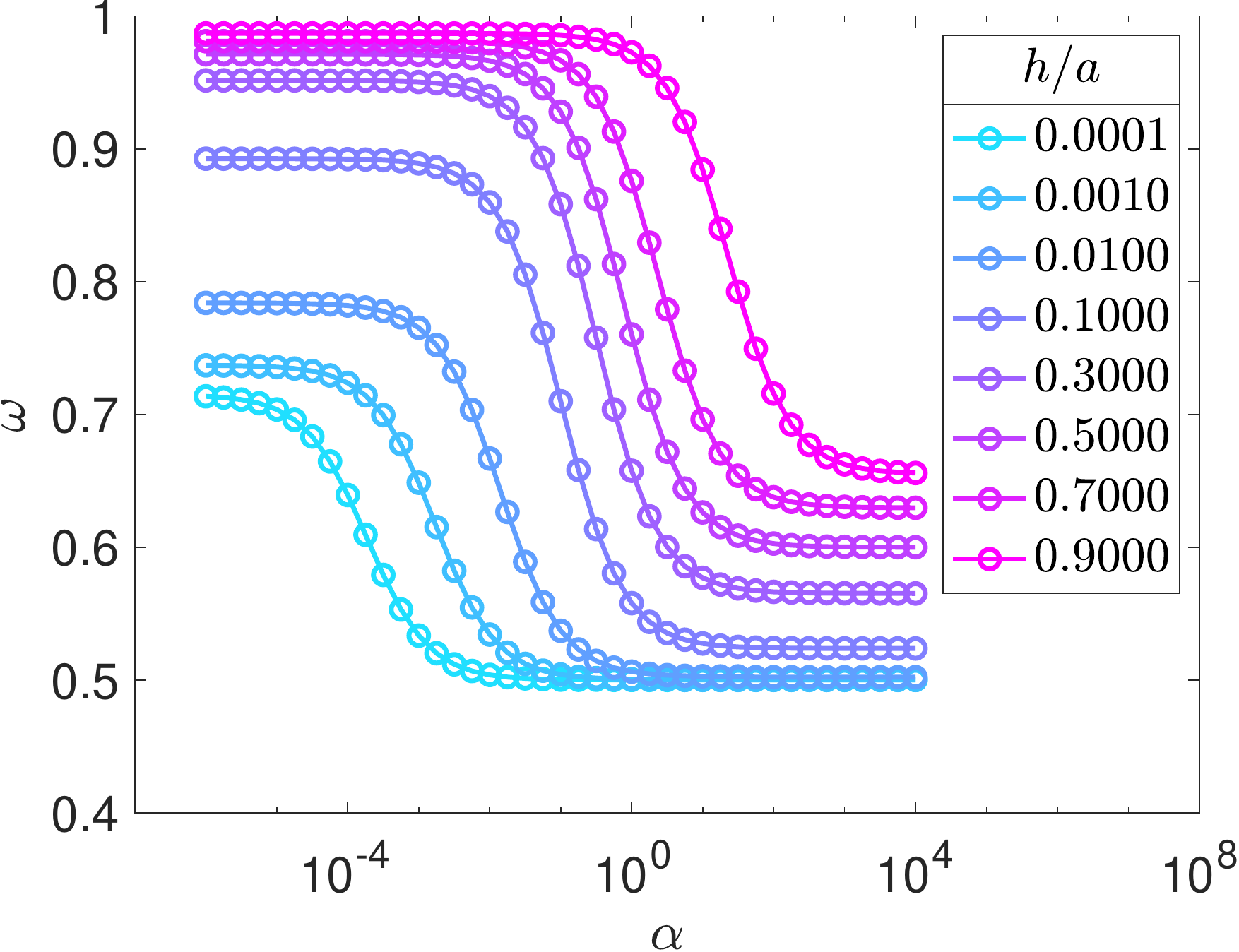}
         \caption{ $h_0/a=1.0$}
         \label{fgr:OmegaAlpha90h01p0}
     \end{subfigure}\\
     \begin{subfigure}[b]{0.3\textwidth}
         \centering
         \includegraphics[width=\textwidth]{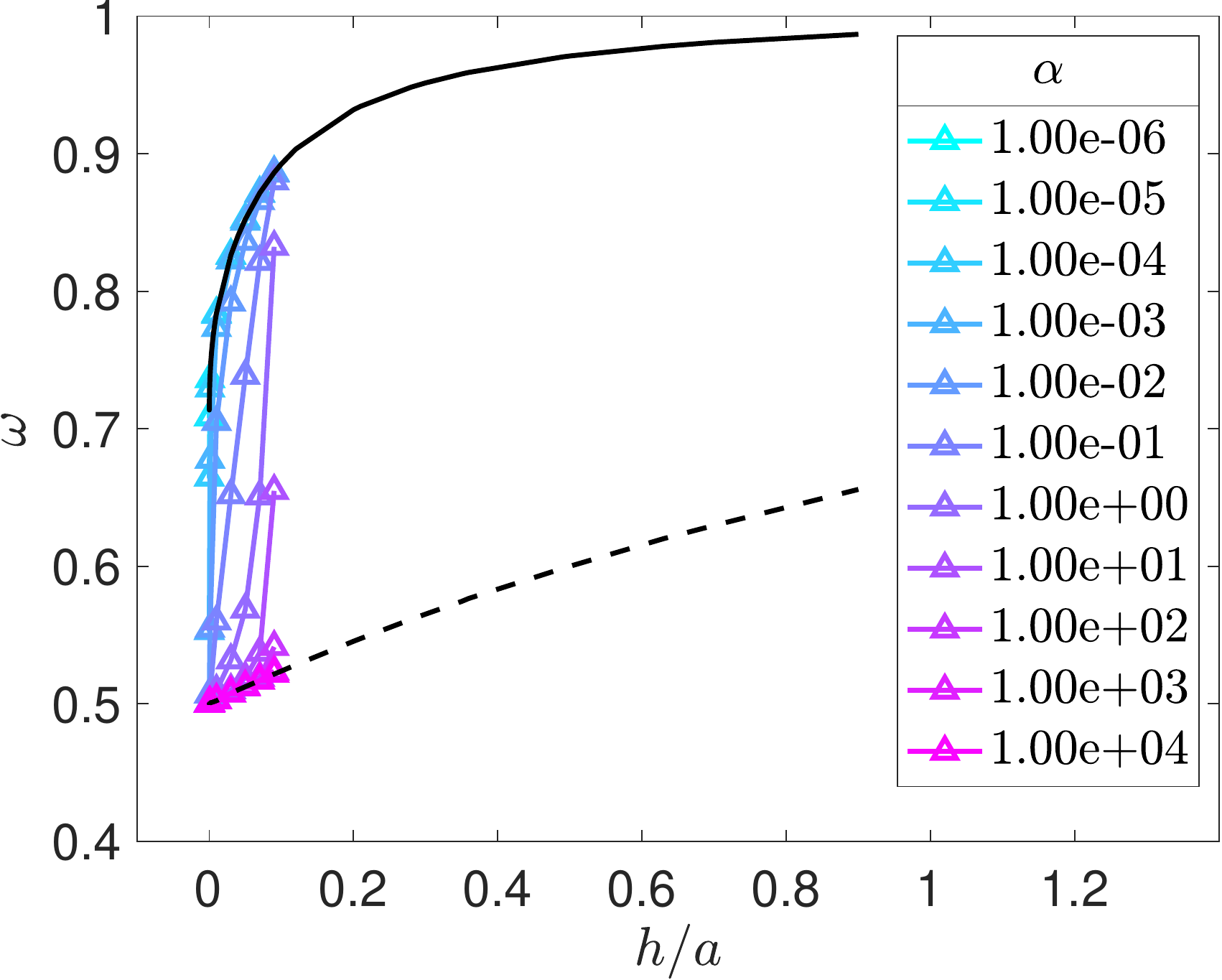}
         \caption{$h_0/a=0.1$ }
         \label{fgr:Omegah90h00p1}
     \end{subfigure}
     \begin{subfigure}[b]{0.3\textwidth}
         \centering
         \includegraphics[width=\textwidth]{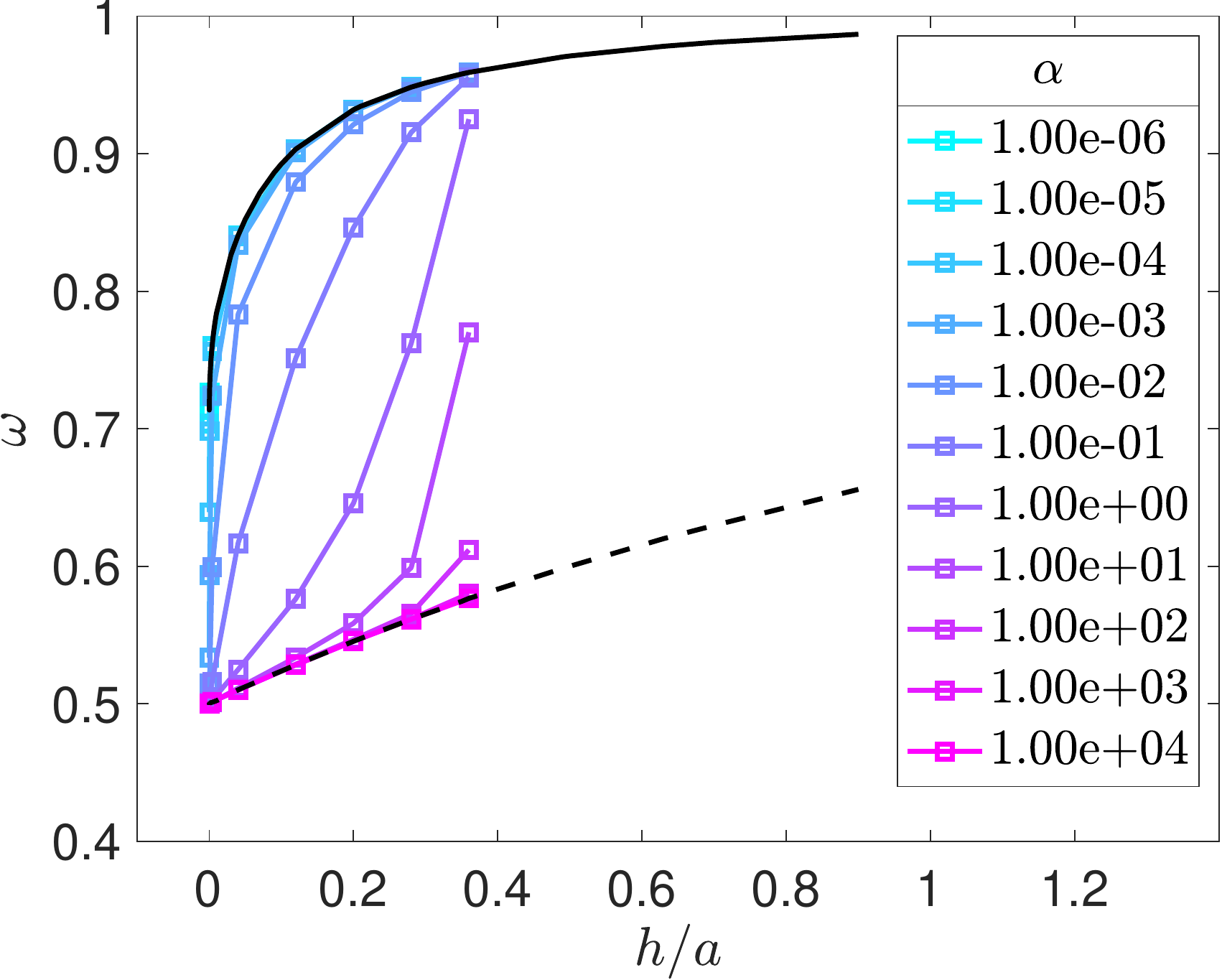}
         \caption{$h_0/a=0.4$ }
         \label{fgr:Omegah90h01p0}
     \end{subfigure}
     \begin{subfigure}[b]{0.3\textwidth}
         \centering
         \includegraphics[width=\textwidth]{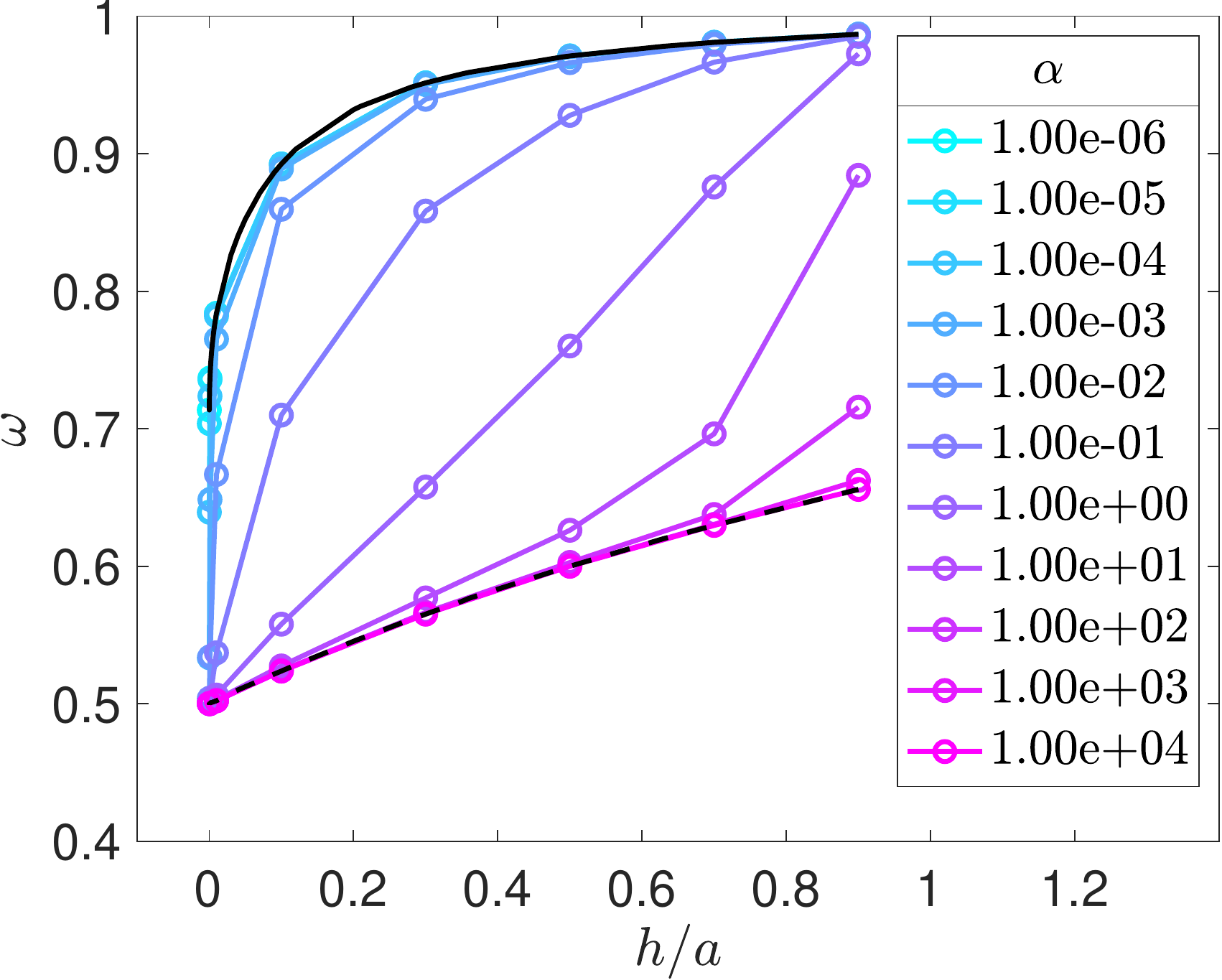}
         \caption{ $h_0/a=1.0$}
         \label{fgr:Omegah90h00p4}
     \end{subfigure}
        \caption{Pair rotation rate $\omega$ as a function of (a) (b) (c) the friction strength $\alpha$ and (d) (e) (f) the separation distance $h/a$ between the particle surfaces for friction ranges of (a) (d) $h_0/a=0.1$, (b) (e) $h_0/a=0.4$ and (c) (f) $h_0/a=1.0$. The particle centers are aligned in the gradient direction ($\theta=90^{\circ}$). The black solid and dashed lines correspond to smooth particle limit and large friction limit respectively. 
        }
        \label{fgr:theta90h00p1h1p0}
\end{figure}

Similar set of curves for different values of the asperity range: $h_0/a = 0.1, 0.4$ and $1.0$ are plotted in figure \ref{fgr:theta90h00p1h1p0}. The $\omega$ versus $\alpha$ and $\omega$ versus $h/a$ curves for all $h_0/a$ values qualitatively follow the same trend as observed for $h_0/a=0.1$. For nearly touching particles: $h/a$ $\ll$ $h_0/a$ the first term of the frictional contribution to the resistance tensor scaling as: $\alpha a/h$ (eq. \ref{frictionFunction}), dominates, and it is independent of the friction range $h_0/a$. Hence the $\omega$ versus $\alpha$ curves for $h/a$ $\ll$ $h_0/a$ are also independent of $h_0/a$. For larger particle separations at the same value of $ \alpha $, the pair rotation rate increases with an increase in $h_0/a$.  This is a result of the transition to the point where the asperities modeled by the additional friction no long overlap.  At the edge of that range, $h/a > h_0/a$, we find the particles move as though they were smooth particles.  The smooth particle limit and high friction (large $\alpha$) limits are plotted as solid and dashed lines, respectively, on all the $\omega$ versus $h/a$ plots. For all the values of friction ranges $h_0/a$, with the decrease in particle separation, the pair rotation rate follows the smooth particle curve for $h/a>h_0/a$. At $h/a\approx h_0/a$, the $\omega$ versus $h/a$ curves deviate from the smooth particle curve. The deviation is sharper for cases where the frictional coupling is stronger.  Eventually, the interparticle separation is close enough that the rotation rate falls on to the curve corresponding to the large $\alpha$ limit.  The particle separation, $ h/a $, at which this transition occurs depends on $ h_0/a $ and $\alpha $. 

As we will discuss later, experimental observation of pair particle trajectories and measurements of the pair rotation rate for different interparticle separation might be mapped onto these figures in order to identify the effective model parameters $ \alpha $ and $ h_0/a $ corresponding to the particles in the experiment.  However, an analytical map describing the curves in figures \ref{fgr:OmegaAlpha90} and \ref{fgr:Omegah90} as well as at other inter-particle separations and orientations is needed to aid in this process.  Here, we discuss a simple rational function model capable of explaining the observed trajectories in these Stokesian Dynamics simulations.

\begin{figure}
 \centering
 \includegraphics[height=6cm]{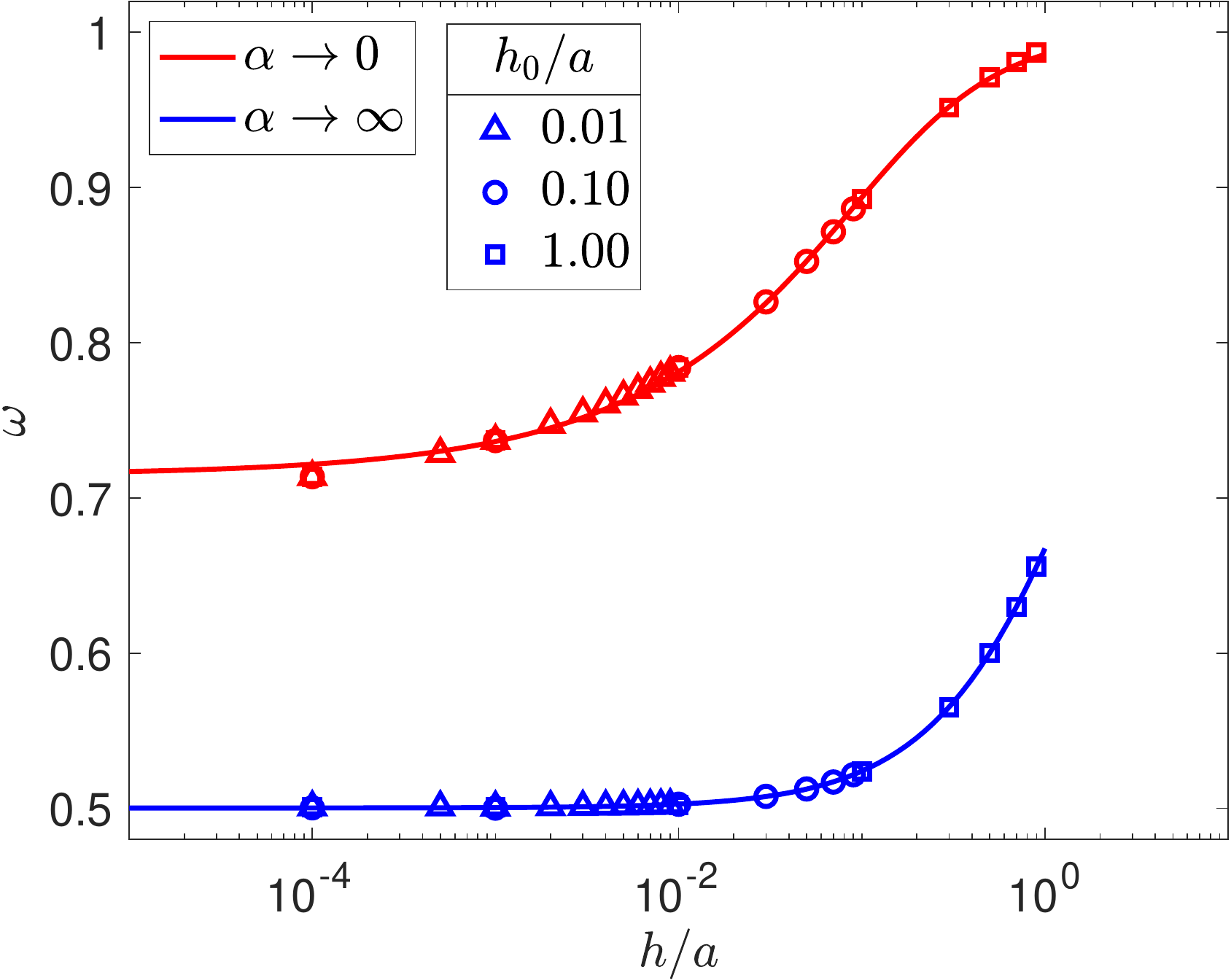}
 \caption{Pair rotation rate $\omega$ as a function of particle separation along the $\theta=90^{\circ}$ line for the two limiting cases of $\alpha$: smooth particle lubrication case ($\alpha\to 0$) and the large friction case ($\alpha\to\infty$), and for the friction range values of $h_0/a=0.01, 0.1$ and $1.0$.}
 \label{fgr:OmegahAlpha0inf}
\end{figure}

First, to find a model that describes the function $\omega(\hat{h}_0,\alpha, \hat{h}, \theta)$ data where $\hat{h} = h/a$ and $\hat{h}_0=h_0/a$, one needs to find combination of variables to nondimentionalize the pair rotation rate. Because of it's sigmoidal character, we consider the nondimensional pair rotation rate:
\begin{equation}
    \hat \omega( \hat{h}_0, \alpha, \hat{h}, \theta ) = \frac{ \omega( \hat{h}_0, \alpha, \hat{h}, \theta ) - \omega_\infty( \hat{h}, \theta ) }{\omega_0( \hat{h}, \theta ) - \omega_\infty( \hat{h}, \theta ) },
\end{equation}
where $ \omega_0( \hat{h}, \theta ) $ and $ \omega_\infty( \hat{h}, \theta )$ represent the pair rotation rate in the limits that $ \alpha \rightarrow 0 $ and $ \alpha \rightarrow \infty $. In this section, we will construct a pair rotation rate model for a pair of particles stacked in gradient direction, $\theta=90^\circ$ and in the next section, we will generalize the model for other pair orientations. Figure \ref{fgr:OmegahAlpha0inf} shows $\omega_0(\hat{h})$ and $\omega_\infty(\hat{h})$ for $\theta=90^\circ$ case. Neither of these functions depends on the range of the asperities, $ \hat{h}_0 $, when $ \hat{h} < \hat{h}_0 $. The simulation data for these limiting curves is well interpolated by the rational functions of the form:

\begin{equation}
    \frac{\omega_{\alpha}(\hat{h})-\omega_{\alpha}(0)}{\omega_{\alpha}(\infty)-\omega_{\alpha}(0)} = \left(1+\frac{m}{\hat{h}^{p}}\right)^{-q},
    \label{limitingcurves}
\end{equation}
where the values of coefficients $(m, p, q)$ obtained using non-linear regression are $(0.133, 1.202, 0.411)$ and $(2.086, 1.023, 0.972)$ for $\alpha\to0$ and $\alpha\to\infty$ cases respectively. The root mean squared error for the fits of the two cases are (0.0105, 0.0006) and the corresponding $R-$squared values are (0.999, 1).

\begin{figure}
 \centering
 \begin{subfigure}[b]{0.4\textwidth}
         \centering
         \includegraphics[width=\textwidth]{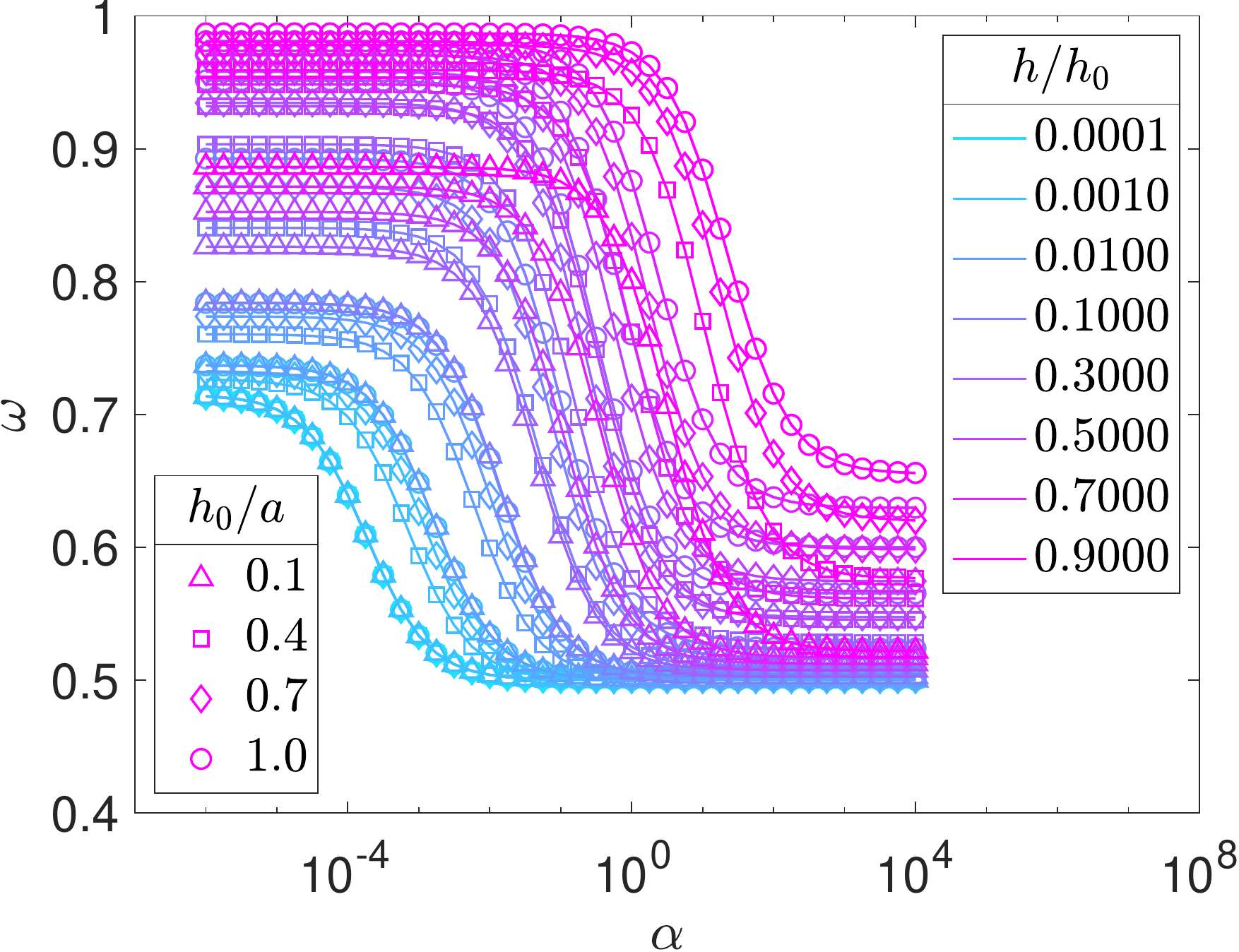}
         \caption{ }
         \label{fig:omegaalpha}
     \end{subfigure}
     \begin{subfigure}[b]{0.4\textwidth}
         \centering
         \includegraphics[width=\textwidth]{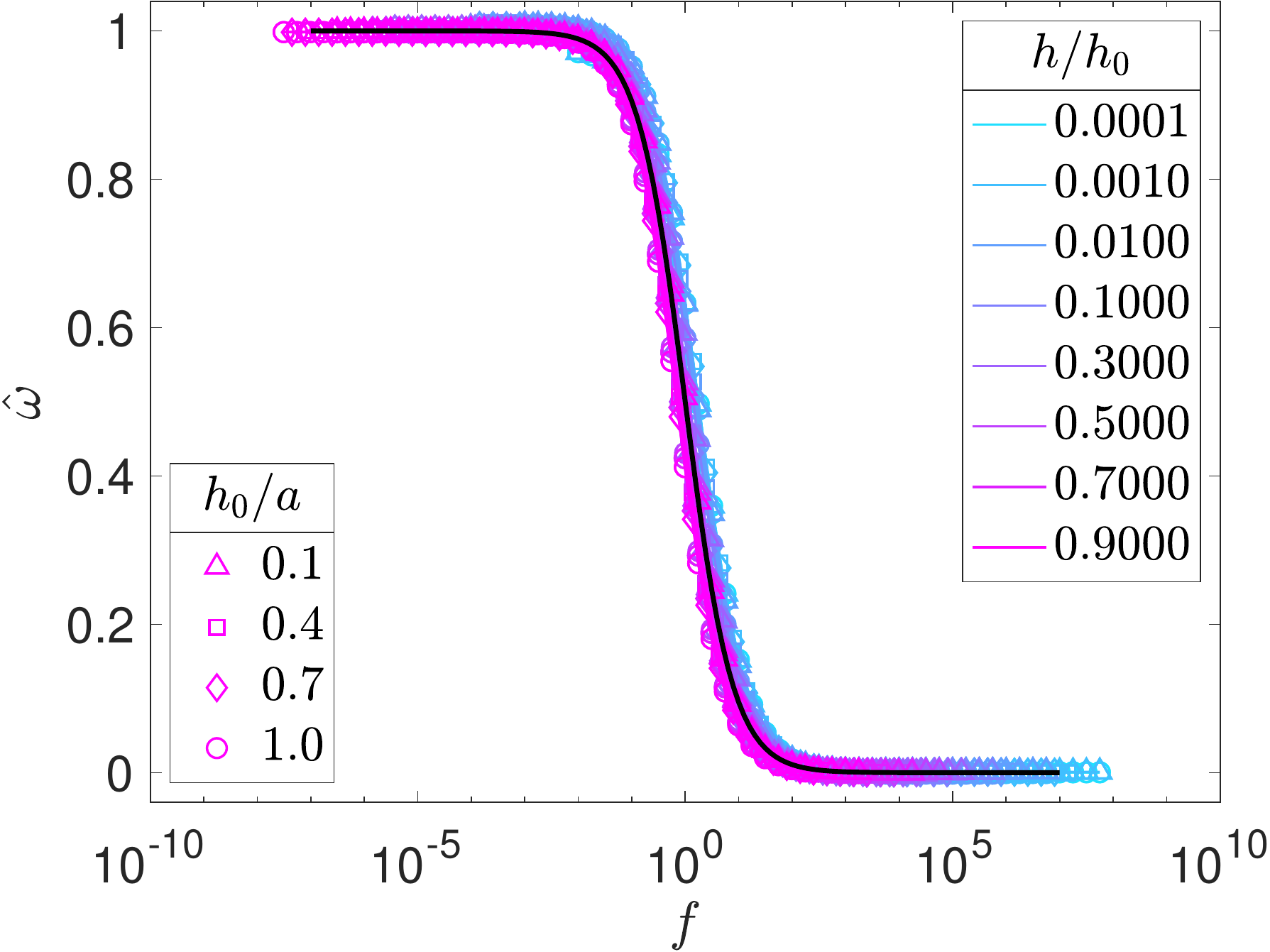}
         \caption{ }
         \label{fig:omegah}
     \end{subfigure}
 \caption{(a) Pair rotation rate, $\omega$, as a function of the friction model parameters, $\alpha$ and $h_0/a$ and particle separation, $h/a$, along $\theta=90^{\circ}$ line. (b) Nondimensionalization of pair rotation rate and particle separation collapses all the data on to the master curve, $\hat{\omega}=1/(1+f^{0.98})$.} 
 \label{fgr:OmegaAlpha90Allh0ND}
\end{figure}

Using $\hat{\omega}$ as the nondimensional pair rotation rate and the sliding resistance function $f$ as the nondimensional input parameter results in collapse of all the $\omega$ versus $\alpha$ curves for $h_0/a = 0.1, 0.4, 0.7$ and $1.0$ and for $10^{-4}<h/a<h_0/a$ on to a master sigmoid curve as shown in figure \ref{fgr:OmegaAlpha90Allh0ND}. Performing non-linear regression on the nondimensional $\hat{\omega}$ versus $f$ data yields the equation \ref{omegaND} for the master curve.

\begin{equation}
    \hat{\omega}(\theta=90^\circ) = \frac{1}{1+f^{0.98}}
    \label{omegaND}
\end{equation}

The polynomial $f$, as given in equation \ref{frictionFunction} and used to compute the sliding resistance, was chosen as the nondimensional particle separation in which $h$ is nondimensionalized with $h_0,\alpha$ and $a$. Since all the $\omega$ versus $\alpha$ plots show inverted sigmoid behavior, the two asymptotic values of the sigmoid are required to nondimenisonalize $\omega$ as equation \ref{omegaND}. The asymptotic values of $\omega$ in the smooth particle limit and large $\alpha$ friction limit computed using Stokesian dynamics simulations are plotted on a semi-log plot as shown in figure \ref{fgr:OmegahAlpha0inf}. The $\omega$ for the smooth particle limiting case, $\alpha\to0$, starts with $0.714$ at close particle separations and increases with $h/a$ to asymptotically reach $\omega=1$ at large particle separations which is equal to the shear rate of the background flow. The large $\alpha$ data for different cases with $h_0/a = 0.01, 0.1$ and $1.0$ follow one master curve and the $\omega$ corresponding to large separations will also reach $\omega=1$ as with the $\alpha\to 0$ case but at larger values of $h/a$ and $h_0/a$. Note that the large friction curve , $\alpha\to\infty$, jumps up sharply to the smooth particle curve at $h/a\approx h_0/a$ over very small region of $h/a$ which is not shown in the figure. The two limiting curves for $\theta=90^{\circ}$ orientations are fitted with equations \ref{limitingcurves} using non-linear regression. The pair rotation rate for nearly touching vertically stacked particles is $\omega_{\alpha\to 0}(h/a\to0)=0.714\dot{\gamma}$ for smooth particles and $\omega_{\alpha\to \infty}(h/a\to0) = \dot{\gamma}/2$ for rough particles with very large friction. For vertically stacked smooth or rough particles separated by very large distances in shear flow, the pair rotation rate is $\omega_{\alpha\to 0}(h/a\to \infty)=\omega_{\alpha\to \infty}(h/a\to \infty)=\dot{\gamma}$.

\subsection{Extension of the pair rotation rate model to other pair orientations}

\begin{figure}
     \centering
     \begin{subfigure}[b]{0.4\textwidth}
         \centering
         \includegraphics[width=\textwidth]{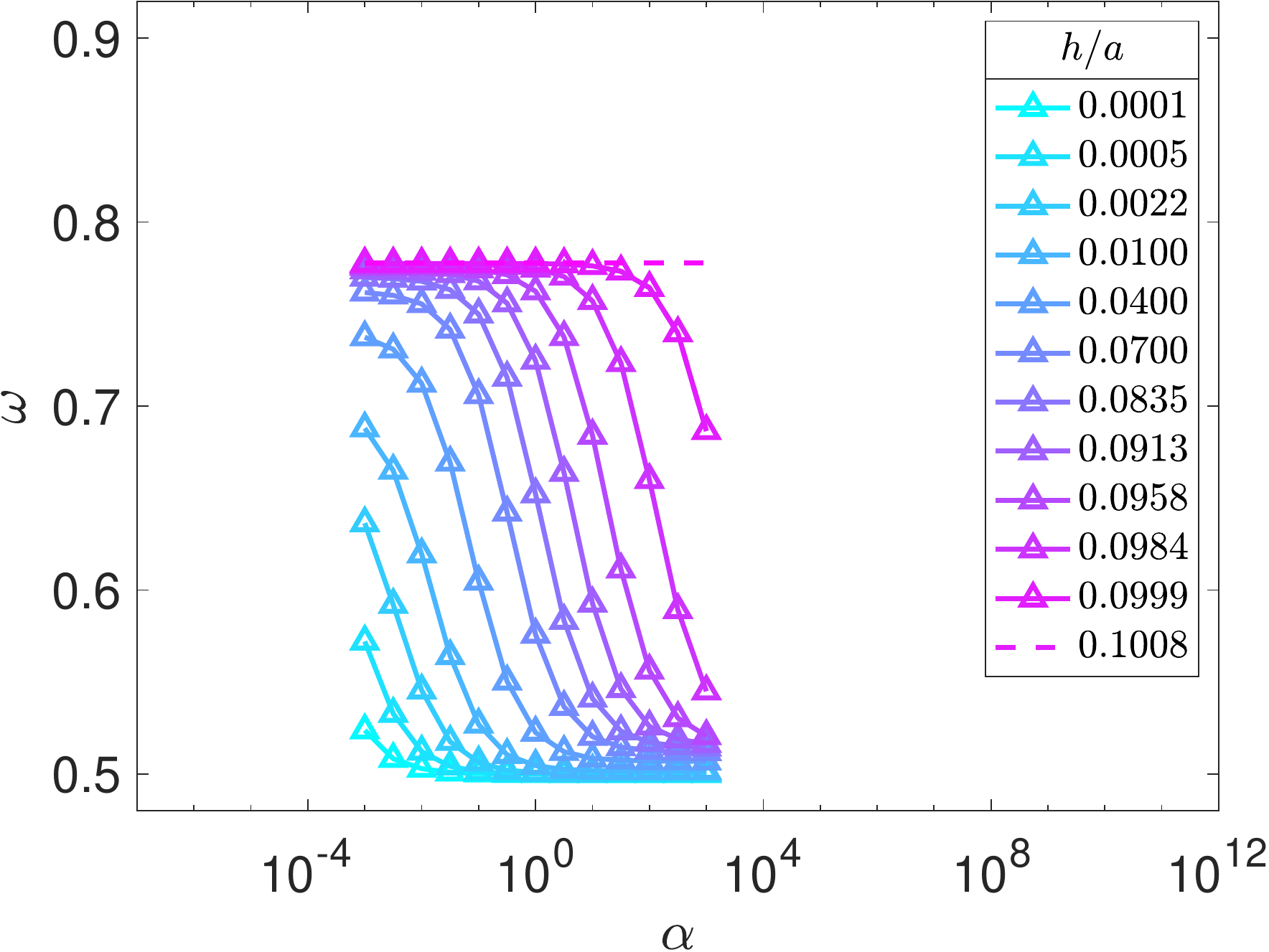}
         \caption{ }
         \label{fig:OmegaAlpha67p5h00p1}
     \end{subfigure}
     \begin{subfigure}[b]{0.4\textwidth}
         \centering
         \includegraphics[width=\textwidth]{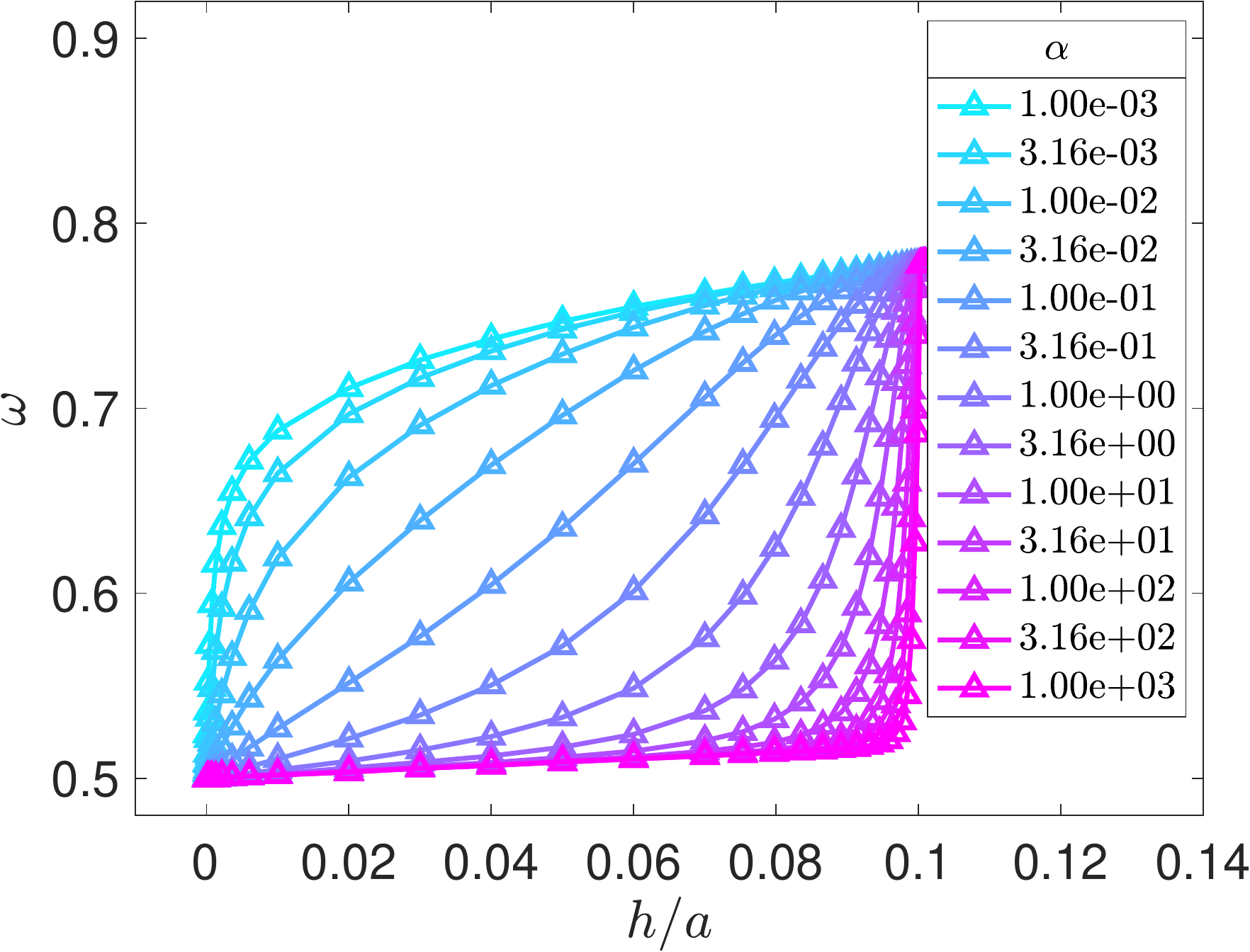}
         \caption{ }
         \label{fig:Omegah67p5h00p1}
     \end{subfigure}
     \begin{subfigure}[b]{0.4\textwidth}
         \centering
         \includegraphics[width=\textwidth]{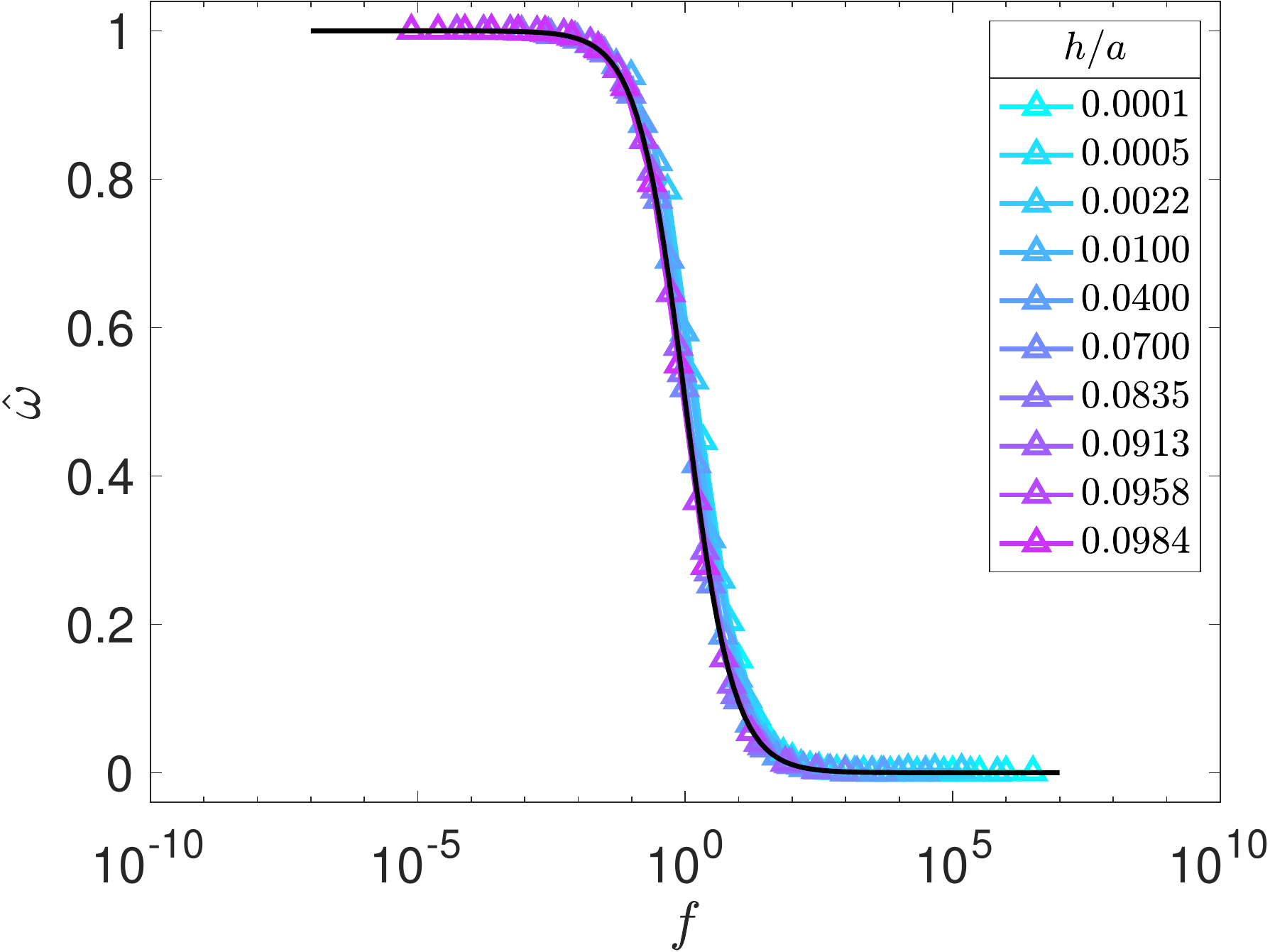}
         \caption{ }
         \label{fig:OmegaAlpha67p5h00p1ND}
     \end{subfigure}
        \caption{Pair rotation rate  at $\theta=67.5^{\circ}$ orientation as a function of (a) the friction strength and (b) the particle separation for the friction range of $h_0/a = 0.1$. Black solid line in (c) is the fit, $\hat{\omega}=1/(1+f^{0.98})$, for the data in (a) and (b) after nondimensionalization.}
        \label{fgr:omega67p5}
\end{figure}

The pair rotation rate as a function of friction strength and particle separation for a pair oriented along $\theta=67.5^{^\circ}$ is plotted in figures \ref{fig:OmegaAlpha67p5h00p1} and \ref{fig:Omegah67p5h00p1} for a friction range of $h_0/a = 0.1$. The curves are qualitatively similar to those for $\theta = 90^{\circ}$ orientation case as shown in figure \ref{fgr:theta90h00p1} but rotation rates are significantly smaller. To scale the curves, the pair rotation rate versus particle separation curves for the limiting cases $\alpha\to0$ and $\alpha\to\infty$ are needed for $\theta=67.5^{^\circ}$ orientation. 
 
\begin{figure}
     \centering
     \begin{subfigure}[b]{0.4\textwidth}
         \centering
         \includegraphics[width=\textwidth]{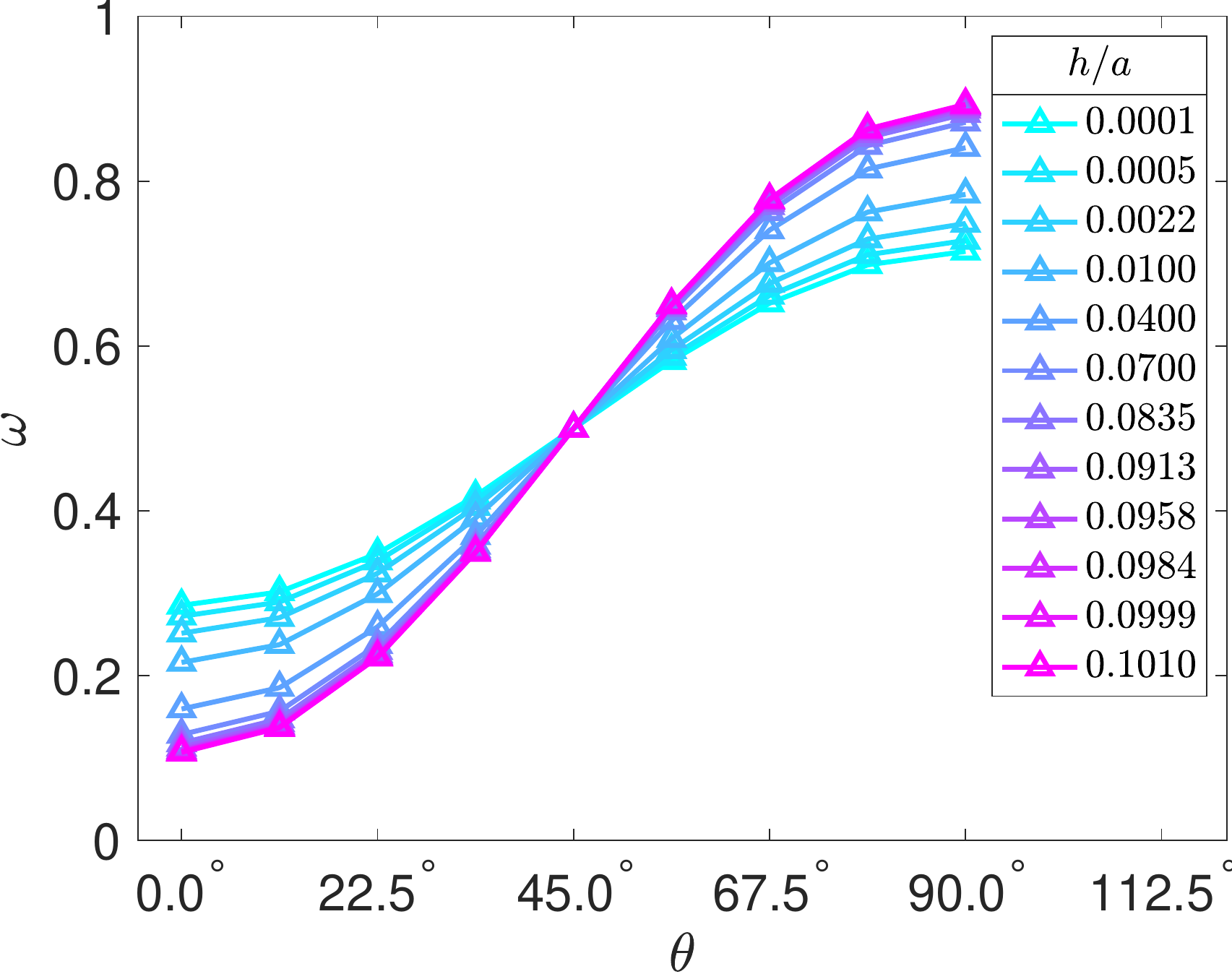}
         \caption{$\alpha=0$.}
         \label{fig:alpha0}
     \end{subfigure}
     \begin{subfigure}[b]{0.4\textwidth}
         \centering
         \includegraphics[width=\textwidth]{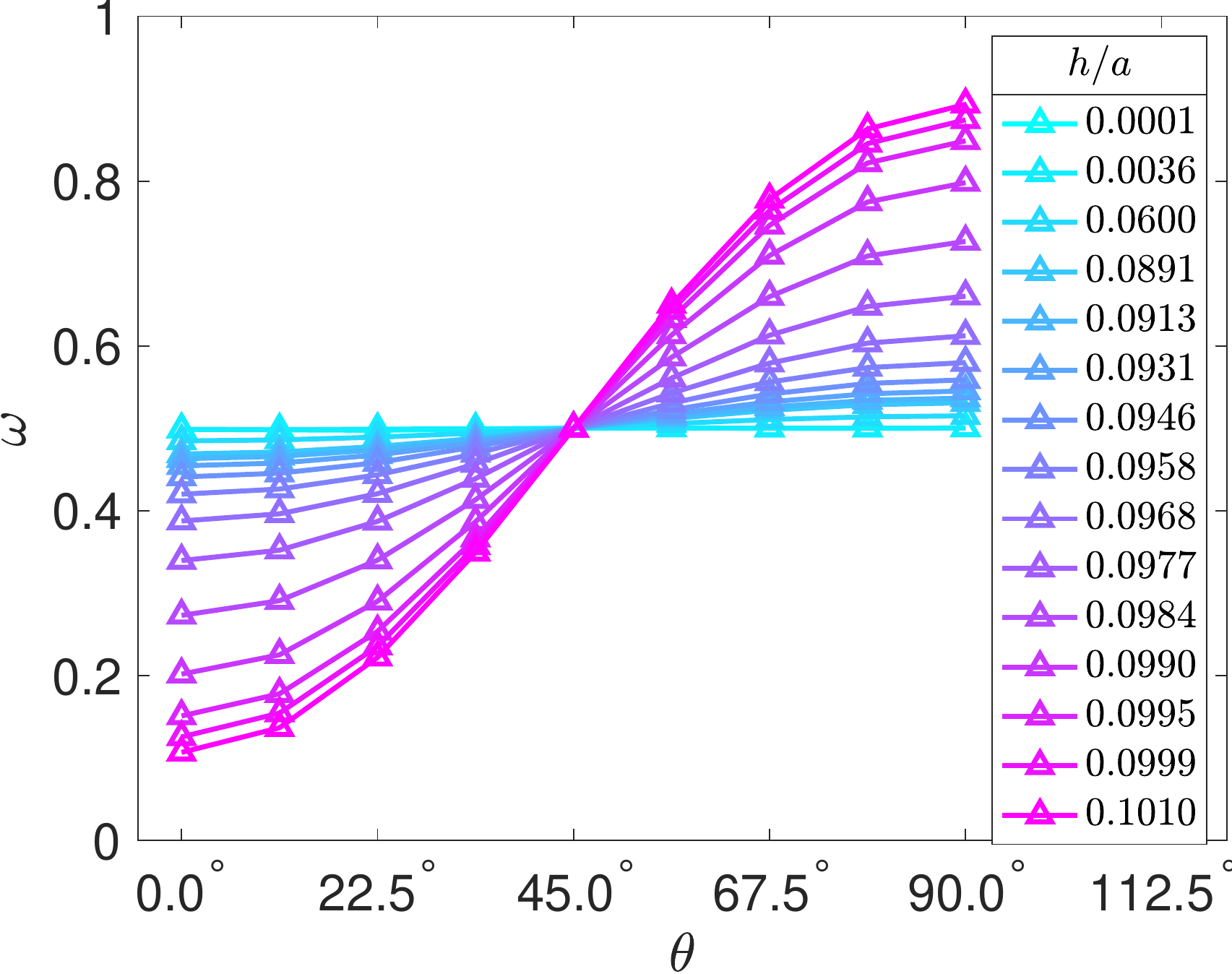}
         \caption{$\alpha=100$}
         \label{fig:alpha100}
     \end{subfigure}
        \caption{Rotation rate of pair particles at different orientations in the compression quadrant for (a) smooth particle case and (b) rough particle case with $\alpha=100$ and $h_0/a = 0.1$. The particle separation distance is varied between $h/a=1e-4$ to $h/a = 0.101$.}
        \label{fgr:OmegaThetanofirc100Alpha}
\end{figure}

The pair rotation rate $\omega$ at different orientation angles in the compression quadrant of the shear flow are plotted for various particle separation distances in figures \ref{fig:alpha0} and \ref{fig:alpha100} for friction strengths of $\alpha=0$ and $\alpha=100$ respectively. The plots show that $\omega$ for orientations along the compression axis ($\theta=45^{\circ}$) is independent of particle separation $h/a$ and friction strength $\alpha$ and equal to $\omega(\theta=45^{\circ})=\dot{\gamma}/2$. This should hold true for all the friction ranges $h_0/a$ as well. The $\omega$ versus $h/a$ curves for $\theta=90^{\circ}$ orientations are already known as shown in figure \ref{fgr:OmegahAlpha0inf} and given by equations \ref{limitingcurves} for $\alpha=0$ and $\alpha\to\infty$. As figures \ref{fig:alpha0} and \ref{fig:alpha100} indicate that the $\omega$ versus $\theta$ behavior is sinusoidal for $0<h/a<h_0/a$, knowing the $\theta=0^{\circ}$ and $\theta=90^{\circ}$ behavior for $\alpha=0$ and $\alpha\to\infty$ is sufficient to model the limiting $\omega$ versus $h/a$ curves for all orientations and the model is given by equation \ref{limitingcurvestheta}.

\begin{equation}
    \frac{\omega_{0}(\theta,h)-\omega(45^{\circ})}{\omega_{0}(90^{\circ},h)-\omega(45^{\circ})} =\frac{\omega_{\infty}(\theta,h)-\omega(45^{\circ})}{\omega_{\infty}(90^{\circ},h)-\omega(45^{\circ})}= sin(2\theta-90^{\circ}) 
    \label{limitingcurvestheta}
\end{equation}

Using the model for limiting curves to nondimensionalize $\omega(h, \alpha)$ data for $\theta = 67.5^{\circ}$ orientation results in collapse of data on to a master curve as shown in figure \ref{fig:OmegaAlpha67p5h00p1ND}. The fit of the master curve is $\hat{\omega} = 1/(1+f^{0.98})$ which is same as that for $\theta=90^{\circ}$ case. This indicate that the fit can describe pair rotation rate for all orientations. The model for the rotation rate $\omega(\alpha,h_0,\theta,h)$ of rough particles subjected to a linear shear flow of strength $\dot{\gamma}$ is summarized in equation \ref{fricmodel} for any orientation in compression quadrant ($0^{\circ}<=\theta<=90^{\circ}, 0<h/a<h_0/a$) of the flow-gradient plane. The limiting curves $\omega_{\alpha\to0}$ and $\omega_{\alpha\to\infty}$ can be computed using equations \ref{limitingcurvestheta} and \ref{limitingcurves}.

\begin{equation}
    \hat \omega( \hat{h}_0, \alpha, \hat{h}, \theta )=\frac{\omega(\alpha,\hat{h}_0,\theta,\hat{h})-\omega_{\alpha\to \infty}(\theta,\hat{h})}{\omega_{\alpha\to 0}(\theta,\hat{h})-\omega_{\alpha\to \infty}(\theta,\hat{h})} = \frac{1}{1+f(\alpha,h_0,a,h)^{0.98}} \\
    \label{fricmodel}
\end{equation}

\subsection{Proposal to experimentally characterize hydrodynamic friction between particles in flow}

\begin{figure}[ht]
     \centering
     \begin{subfigure}[b]{0.29 \textwidth}
         \centering
         \includegraphics[width=\textwidth]{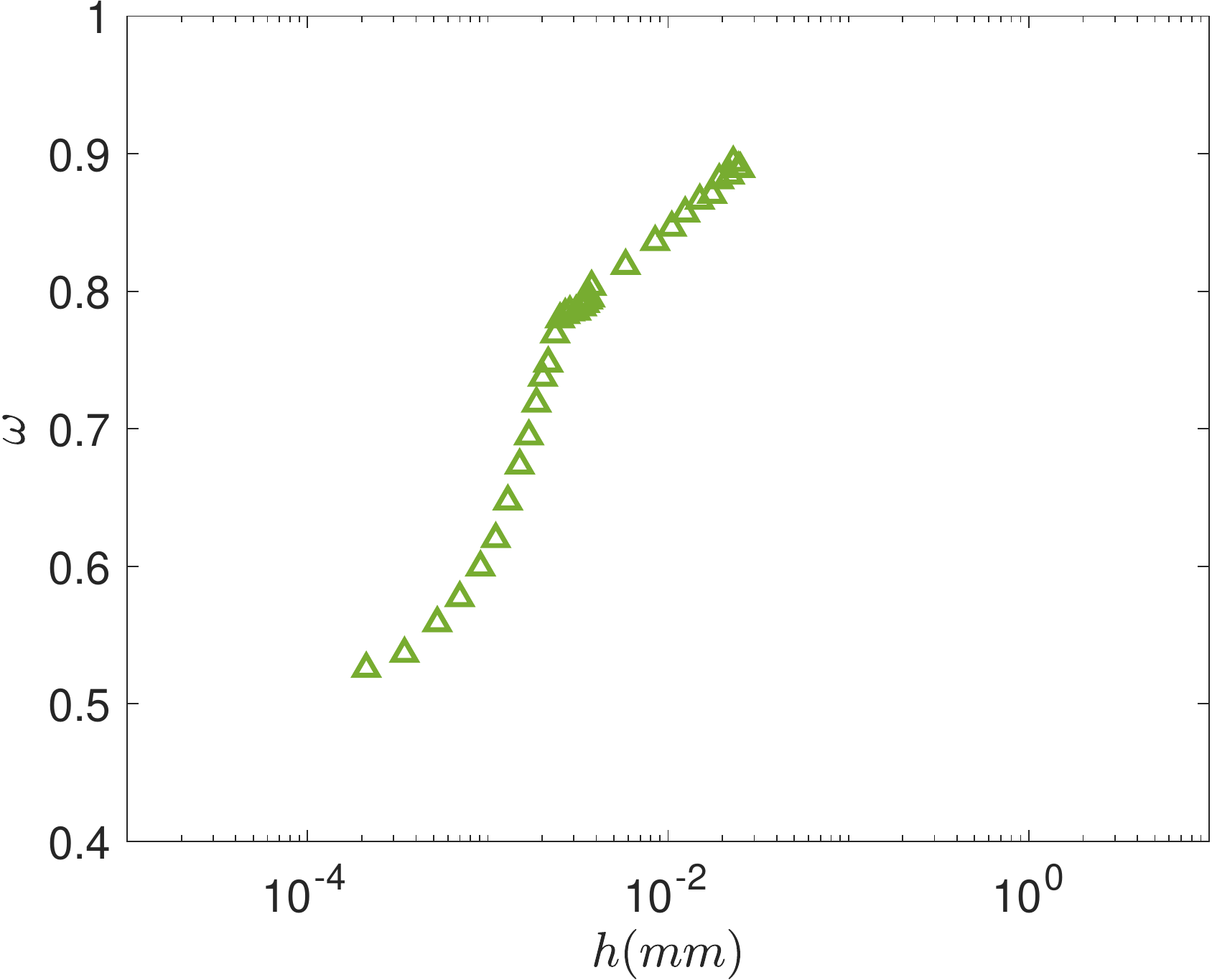}
         \caption{Raw experimental data.}
         \label{fig:exptmm}
     \end{subfigure}
     \hfill
     \begin{subfigure}[b]{0.3\textwidth}
         \centering
         \includegraphics[width=\textwidth]{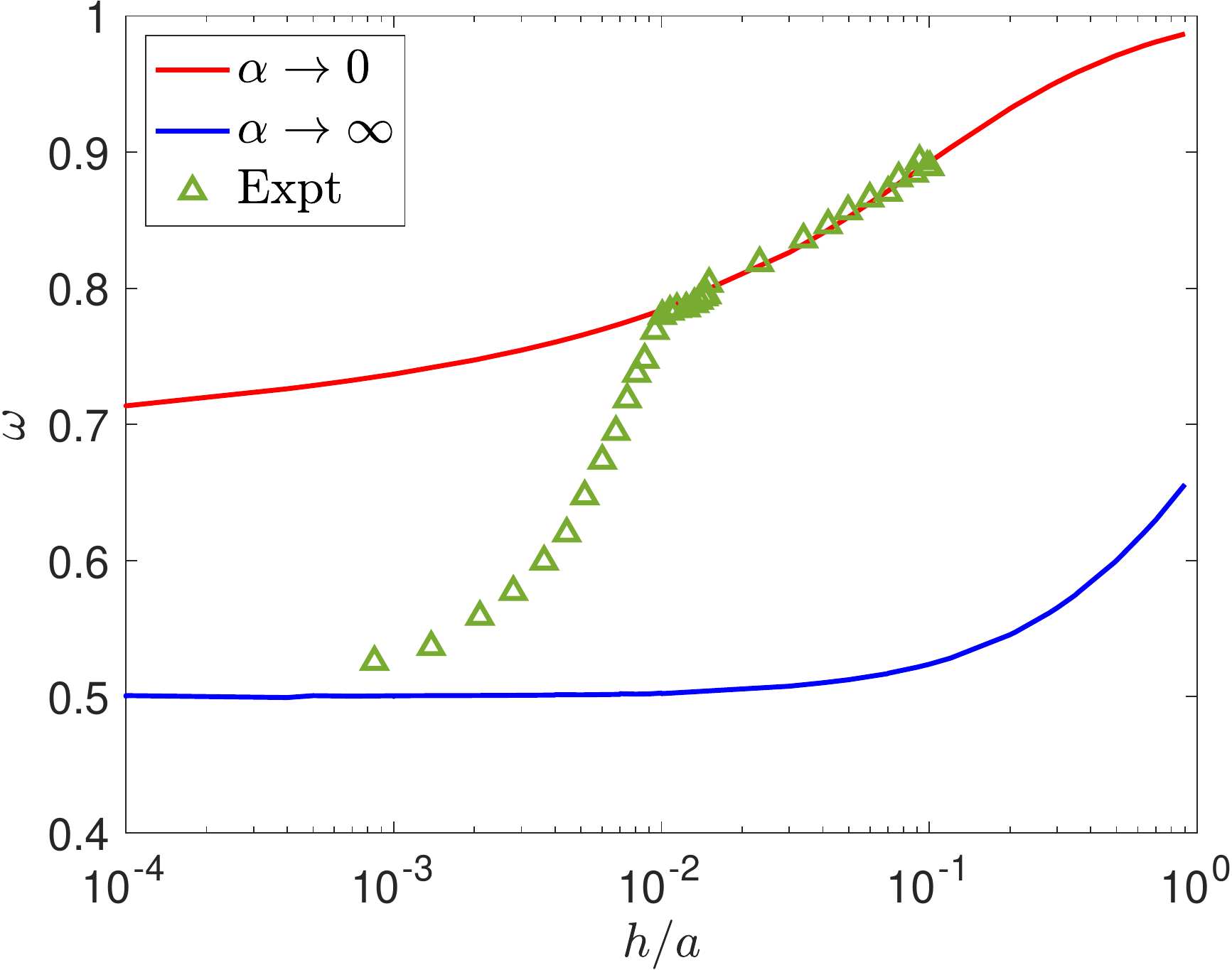}
         \caption{Finding $r_h$ and $h_0$}
         \label{fig:expth0}
     \end{subfigure}
     \hfill
     \begin{subfigure}[b]{0.3\textwidth}
         \centering
         \includegraphics[width=\textwidth]{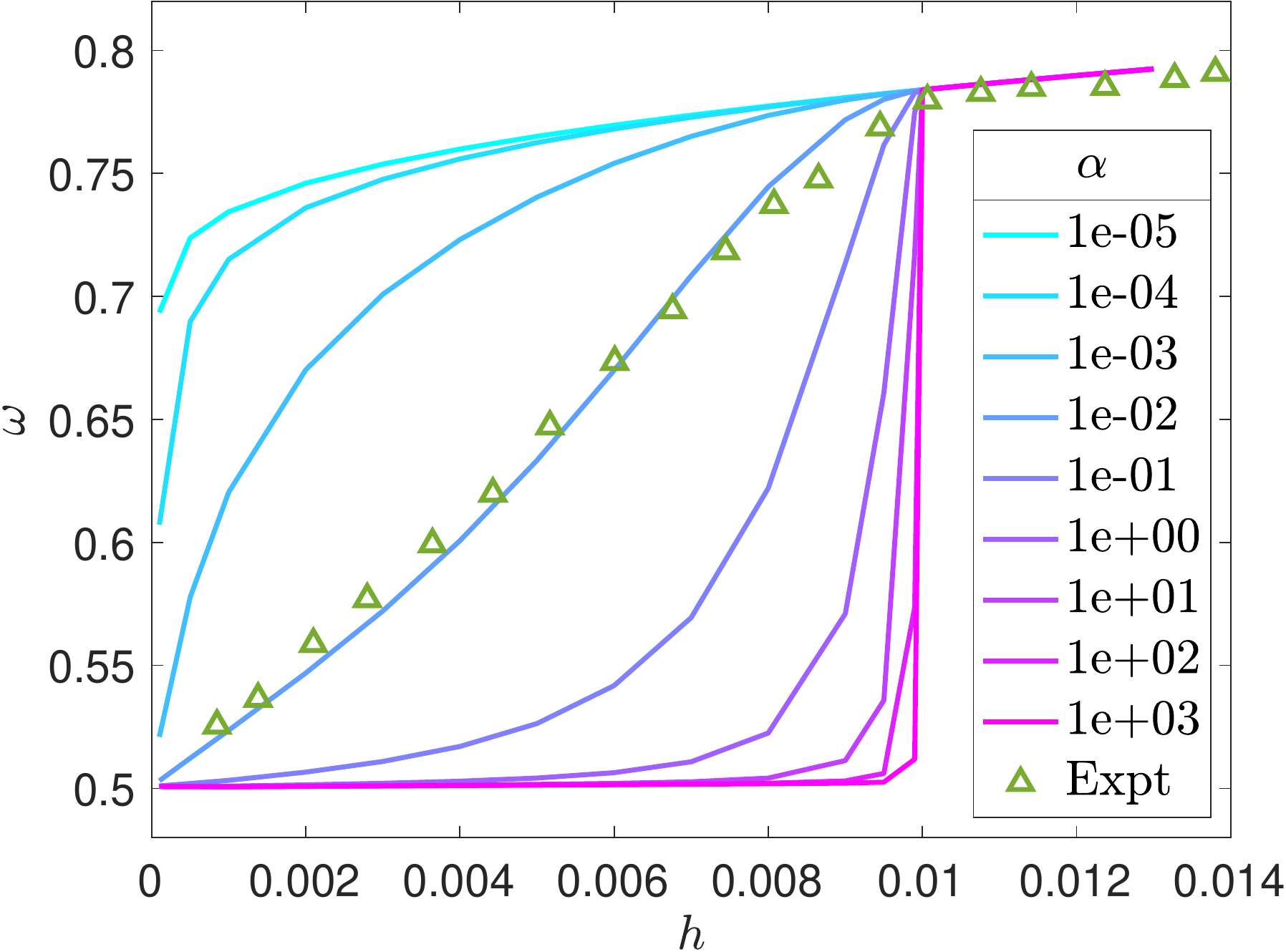}
         \caption{Finding $\alpha$}
         \label{fig:exptalpha}
     \end{subfigure}
        \caption{Data in green triangles is a fictitious experimental data constructed to demonstrate the protocol to find size, friction range and friction strength of rough particles. (a) Experimental pair rotation rate at various particle separations $h$ for particle orientations along the gradient direction. (b) Experimental $h$ is scaled with 0.25 $mm$ to match the smooth particle limit curve generated using the limiting curve model given by equations \ref{limitingcurvestheta} and \ref{limitingcurves} for $\theta = 90^{\circ}$, indicating a particle size $r_h\approx250$ $\mu m$. The experimental curve deviates from the $\alpha \to 0$ limiting curve around $h/a=0.01$ indicating a friction range of $h_0\approx0.01r_h$. (c) $\omega$ versus $h$ curves are computed for $h_0/a=0.01$ and various values of $\alpha$ using the pair rotation model given by equation \ref{fricmodel}. The experimental data matches the model predictions for $\alpha\approx0.01$ indicating a friction strength of $0.01$.}
        \label{fgr:exptdemo}
\end{figure}

The pair rotation rate model in equation \ref{fricmodel} can be used to experimentally characterize the hydrodynamic friction between particles in suspension flows as follows. Subject a neutrally buoyant dilute suspension of rough particles to simple shear flow at very small particle scale Reynolds numbers $Re_p = 2a^2\dot{\gamma}/\nu\ll1$ in an experimental set-up, where $\nu$ is the kinematic viscosity of the fluid. Image the particle positions with time in a flow-gradient plane of thickness in the vorticity direction much smaller than the particle size. Identify all the pair interactions and using their position and time data,  compute the relative velocity normal to the particle center-line and particle separation distance $h$ between the two particles of the pair when the particle center line aligns in the gradient direction ($\theta=90^{\circ}$). Compute the experimental pair rotation rate $\omega_{expt}$ as a function of $h$ by taking the ratio of the tangential relative velocity and particle center to center distance. Using the pair rotation rate model given in equation \ref{limitingcurvestheta}, generate $\alpha\to0$ and $\alpha\to\infty$ limiting curves for $\theta=90^{\circ}$ which are plotted in figure \ref{fig:expth0}. The average hydrodynamic radius $r_h$ of the rough particles can be found by matching the experimental $\omega$ versus $h/r_h$ curve with the model generated limiting curve $\omega_{\alpha\to0}$ versus $h/a$ over the separation range $h_0/a<h/a<1$. The $h/a$ below which the experimental curve deviates from the smooth particle limiting case gives the estimation of the friction range $h_0/a$. Using this $h_0/a$, the model generated limiting curves and the pair rotation rate model from equation \ref{fricmodel}, one can generate $\omega$ versus $h/a$ curves for various values of $\alpha$ and $\theta=90^{\circ}$ which are shown in figure \ref{fig:exptalpha}. Predict the $\alpha$ by matching the experimental $\omega$ versus $h/a$ curve with the model generated curves. A fictitious experimental data is used to demonstrate in figure \ref{fgr:exptdemo} the method to find the effective size, friction range and friction strength of a given pair of rough particles. The method yields same results for any pair orientation $0^{\circ}<\theta<=90^{\circ}$ except for $\theta=45^{\circ}$ as it is the trivial case. However, $\theta=90^{\circ}$ orientation is the relatively easier case to experimentally measure the tangential relative velocity and separation distance between the pair of particles.  

\subsection{Rheology of frictional suspensions}
To compute the hydrodynamic contribution to the high frequency viscosity of a suspension of rough particles, Stokesian dynamics simulations incorporated with the friction model are performed for 100 independent hard sphere configurations each with 2000 monodisperse spherical particles in a cubic  box corresponding to a given volume fraction between $\phi=0.01$ and $\phi=0.63$. The random hard sphere configurations are generated using an event-driven molecular dynamics algorithm as described in Skoge~\textit{et al.}~\cite{skoge2006packing}. Each of the configuration is sheared athermally at shear rate $\dot{\gamma}$ and the simulation is run for one diffusion time in one discrete time step to compute the hydrodynamic stresslet $S_{xy}$ on each the particle. Lees Edwards boundary conditions are applied in the gradient direction and periodic boundary conditions are applied in the flow and vorticity directions. In the absence of thermal forces on the particles and external potentials between the particles, the particle average stresslet over all the configurations $<S_{xy}>$ is the only contribution to the suspension shear stress. The high frequency shear viscosity of the suspension relative to the solvent contribution $\eta_r$ and the fluctuations in the stresslet $S_{xy}^{'}$ are computed using  equations \ref{highfeqviscSxyfluct}. The effect of the parameters of the friction model: friction range $h_0/a$ and friction strength $\alpha$ on the suspension viscosity and stress fluctuations are explored over a range of particle volume fractions.

\begin{equation}
    \begin{split}
    \eta_r &= 1 + \frac{9}{2}\frac{\phi<S_{xy}>}{6\pi a^3 \eta \dot{\gamma}} \\
    S_{xy}^{'} &= \frac{<S_{xy}^2>-<S_{xy}>^2}{<S_{xy}>}
    \end{split}
    \label{highfeqviscSxyfluct}
\end{equation}
\begin{figure}[ht]
     \centering
     \begin{subfigure}[b]{0.4\textwidth}
         \centering
         \includegraphics[width=\textwidth]{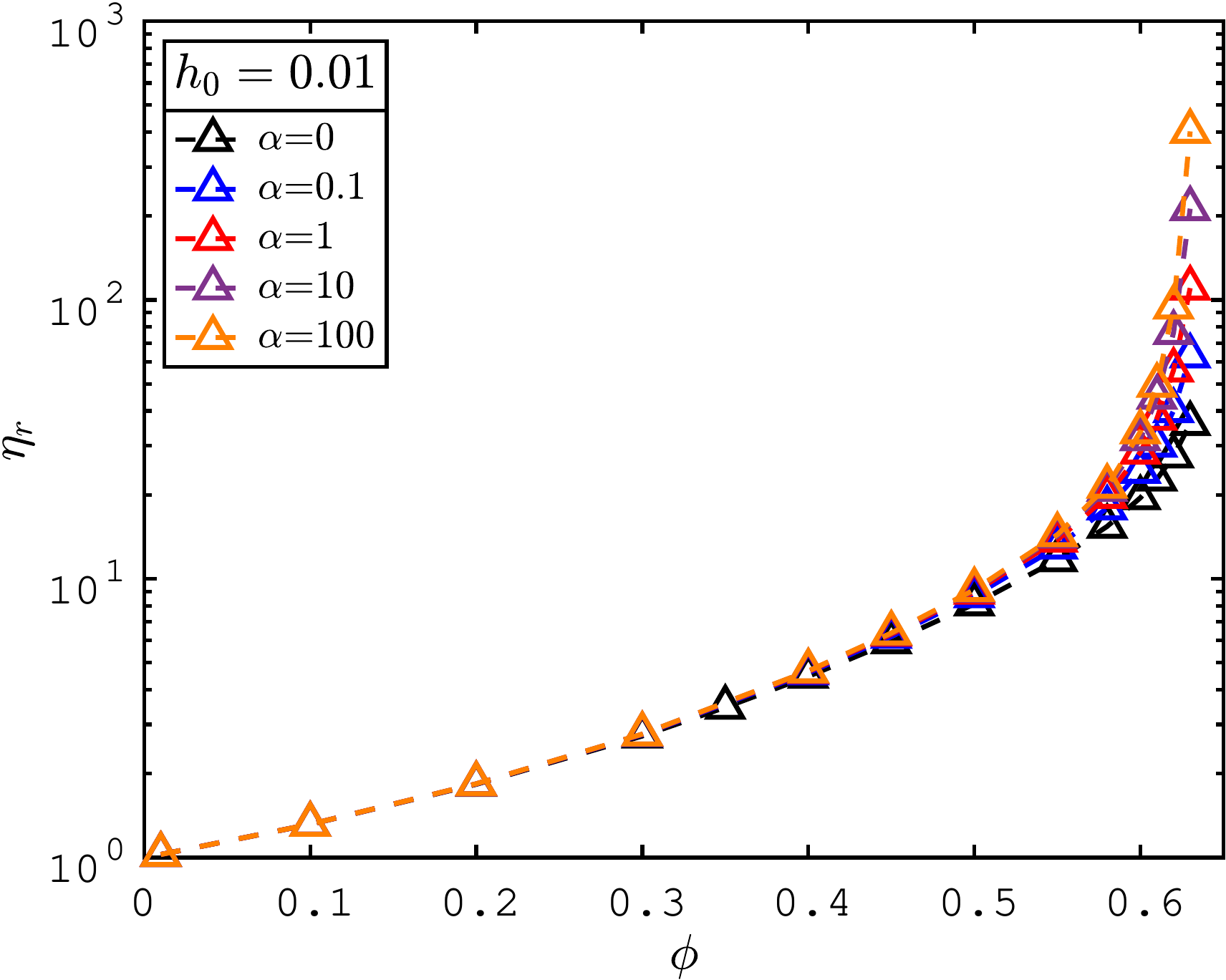}
         \caption{}
         \label{fig:etar_h0_0.01}
     \end{subfigure}
     \begin{subfigure}[b]{0.4\textwidth}
         \centering
         \includegraphics[width=\textwidth]{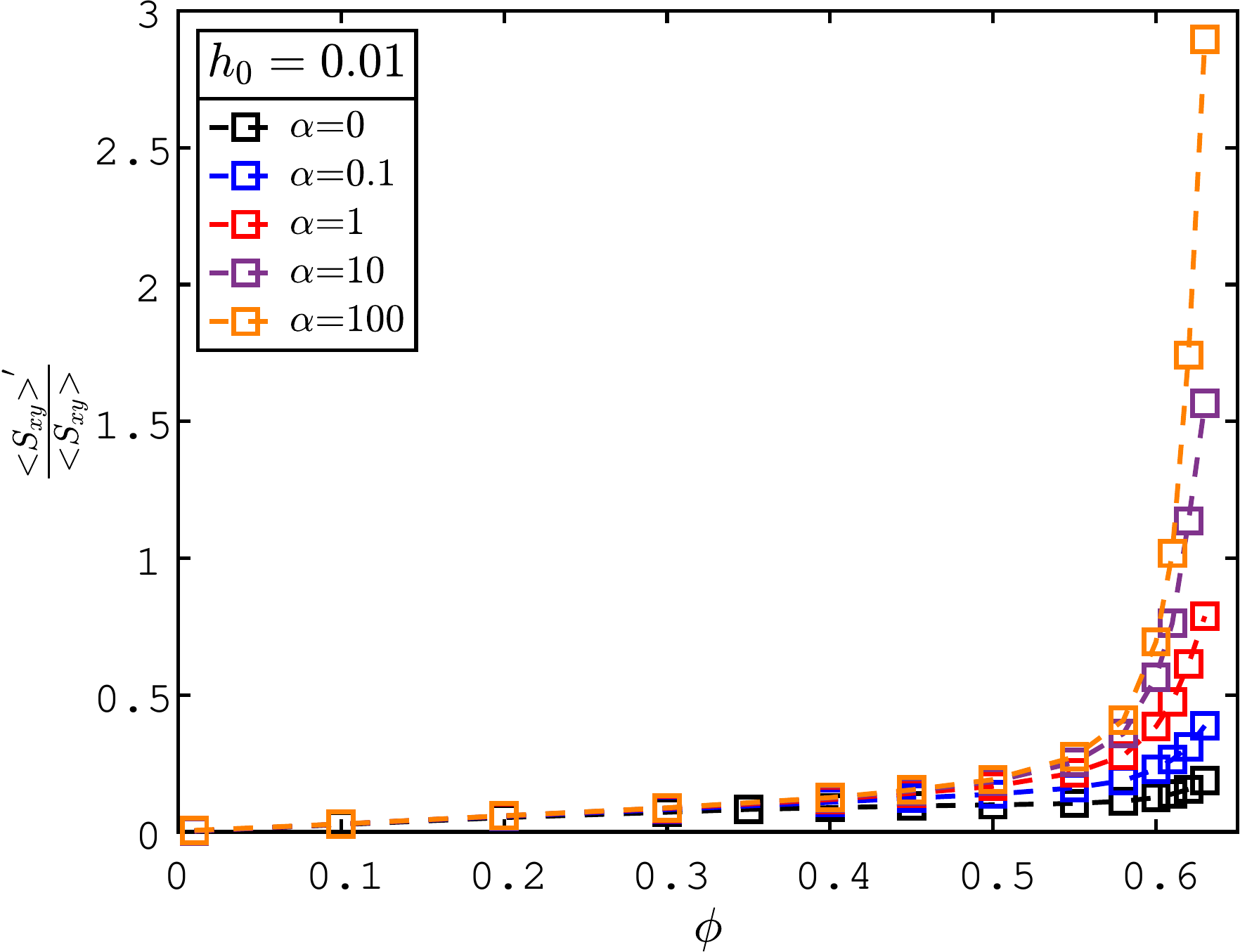}
         \caption{}
         \label{fig:Sxyfluct_h0_0.01}
     \end{subfigure}
     \begin{subfigure}[b]{0.4\textwidth}
         \centering
         \includegraphics[width=\textwidth]{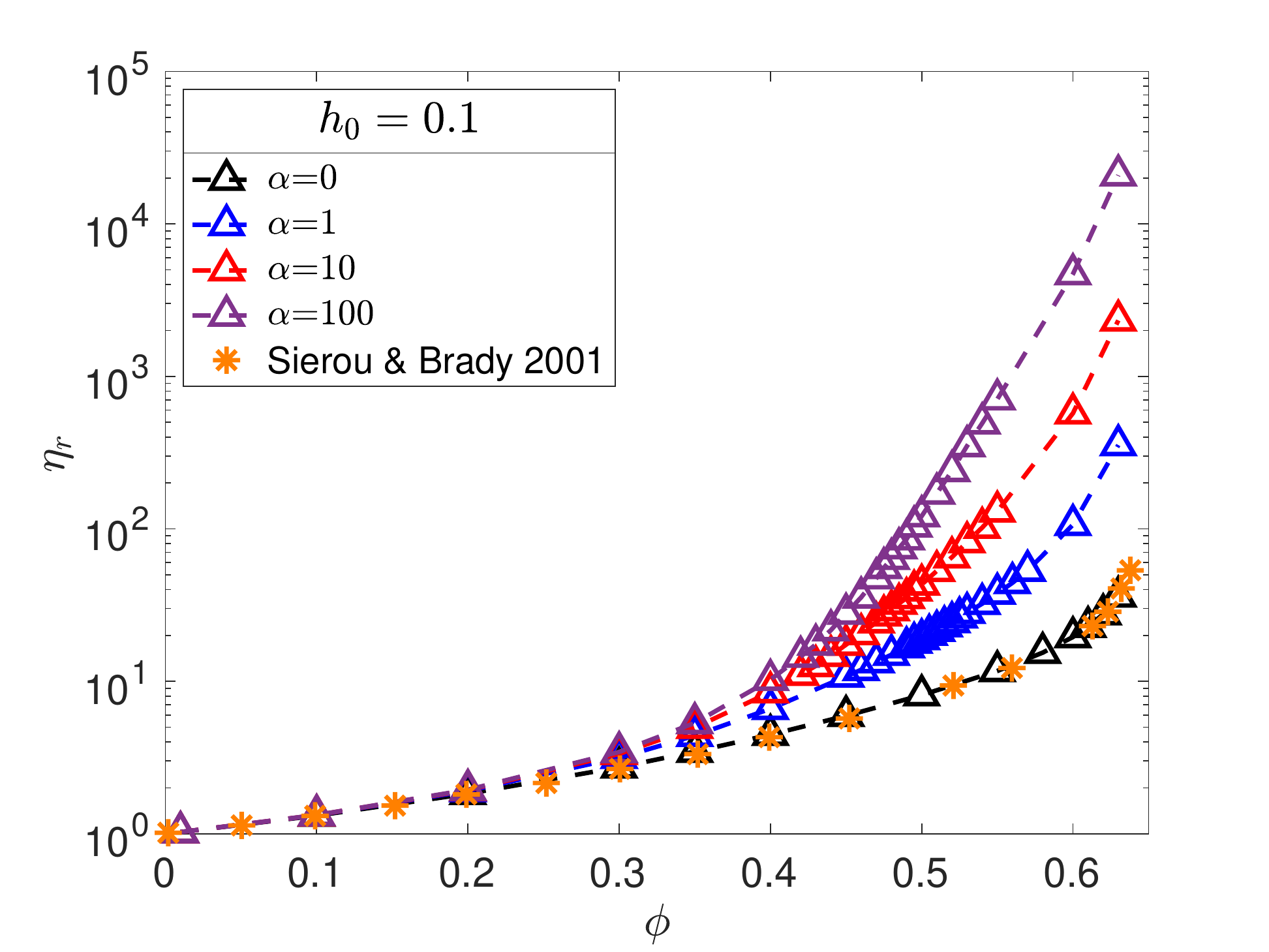}
         \caption{}
         \label{fig:etar_h0_0.1}
     \end{subfigure}
     \begin{subfigure}[b]{0.4\textwidth}
         \centering
         \includegraphics[width=\textwidth]{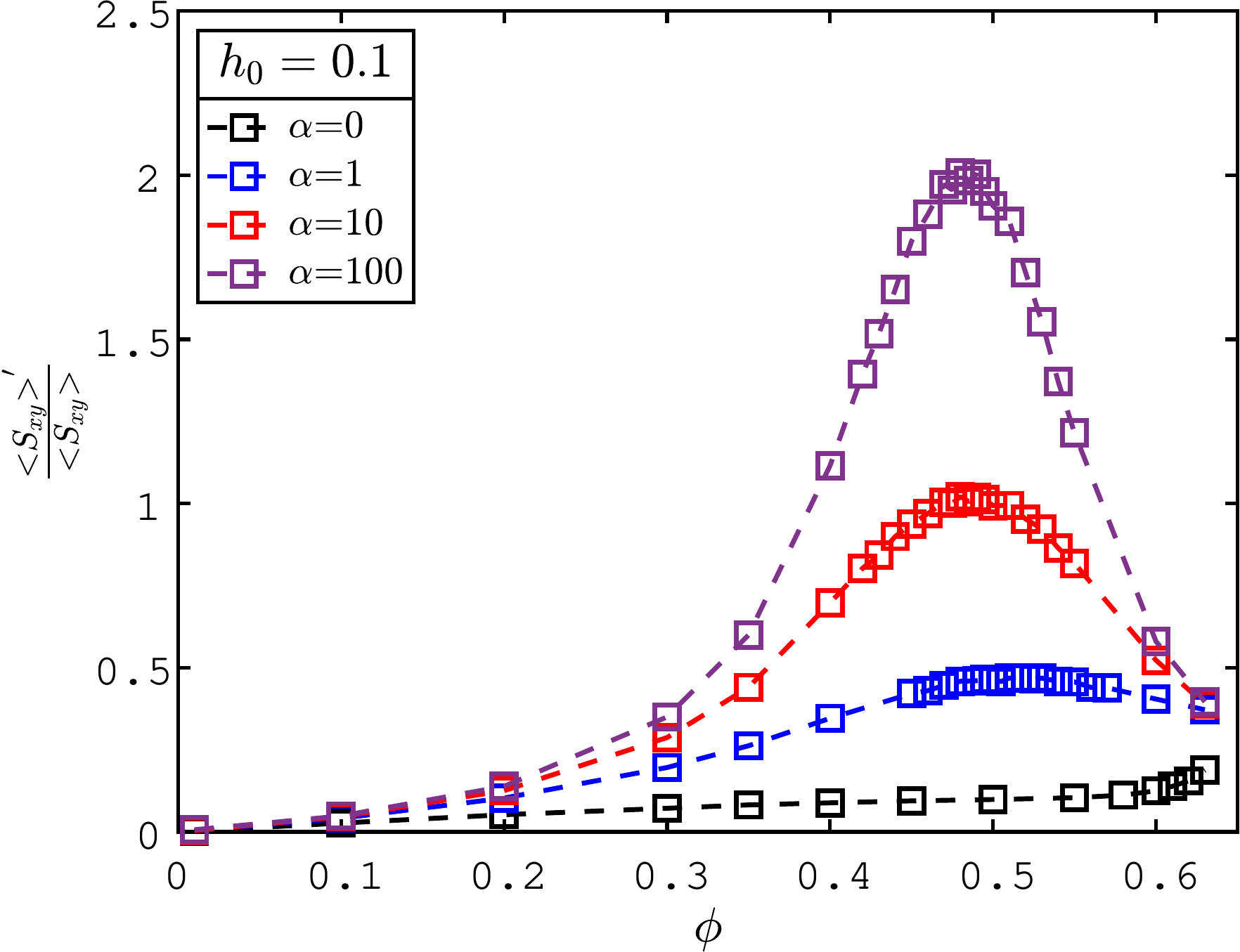}
         \caption{}
         \label{fig:Sxyfluct_h0_0.1}
     \end{subfigure}
        \caption{High frequency shear viscosity  and  normalized stresslet fluctuations of a suspension as a function of particle volume fraction when friction strength is varied between $\alpha=0$ and $\alpha=100$ for two cases of friction range:($a$) and ($b$) $h_0/a=0.01$ and  ($b$) and ($d$) $h_0/a=0.1$.}
        \label{fgr:etarSxyfluctalphavary}
\end{figure}

Figures \ref{fig:etar_h0_0.01} and \ref{fig:etar_h0_0.1} show the effect of the friction strength $\alpha$ on the high frequency suspension viscosity versus particle volume fraction curve for two different values of friction range, $h_0/a=0.01$ and $h_0/a=0.1$. Corresponding influence of $\alpha$ on the stresslet fluctuations normalized with the average stresslet is shown in figures \ref{fig:Sxyfluct_h0_0.01} and \ref{fig:Sxyfluct_h0_0.1}. Note that the fluctuations in stresslet directly correspond to the fluctuations in suspension stress here. The $\alpha=0$ case in figures \ref{fig:etar_h0_0.01} and \ref{fig:etar_h0_0.1} corresponds to the smooth particle Stokesian dynamics which agrees well with the accelerated Stokesian dynamic simulations of Sierou and Brady~\cite{sierou2001accelerated} as shown in figure \ref{fig:etar_h0_0.1}. At lower concentrations the high frequency viscosity is independent of the friction strength $\alpha$ and at higher concentrations increase in $\alpha$ results in increased $\eta_r$. The $\phi$ above which the $\alpha$ dependence becomes significant depends strongly on the friction range $h_0/a$. For higher $h_0/a$, the particles experience hydrodynamic frictional coupling at relatively larger average particle separations compared to the case with smaller $h_0/a$. This results in relatively higher viscosity at lower $\phi$ values. At the same particle loading, $\phi=0.63$ for example, when $\alpha$ is increased from $\alpha=0$ to $\alpha=100$ the viscosity increases by one order of magnitude for $h_0/a=0.01$ and three orders of magnitude for $h_0/a=0.1$. 

\begin{figure}[ht]
     \centering
     \begin{subfigure}[b]{0.4\textwidth}
         \centering
         \includegraphics[width=\textwidth]{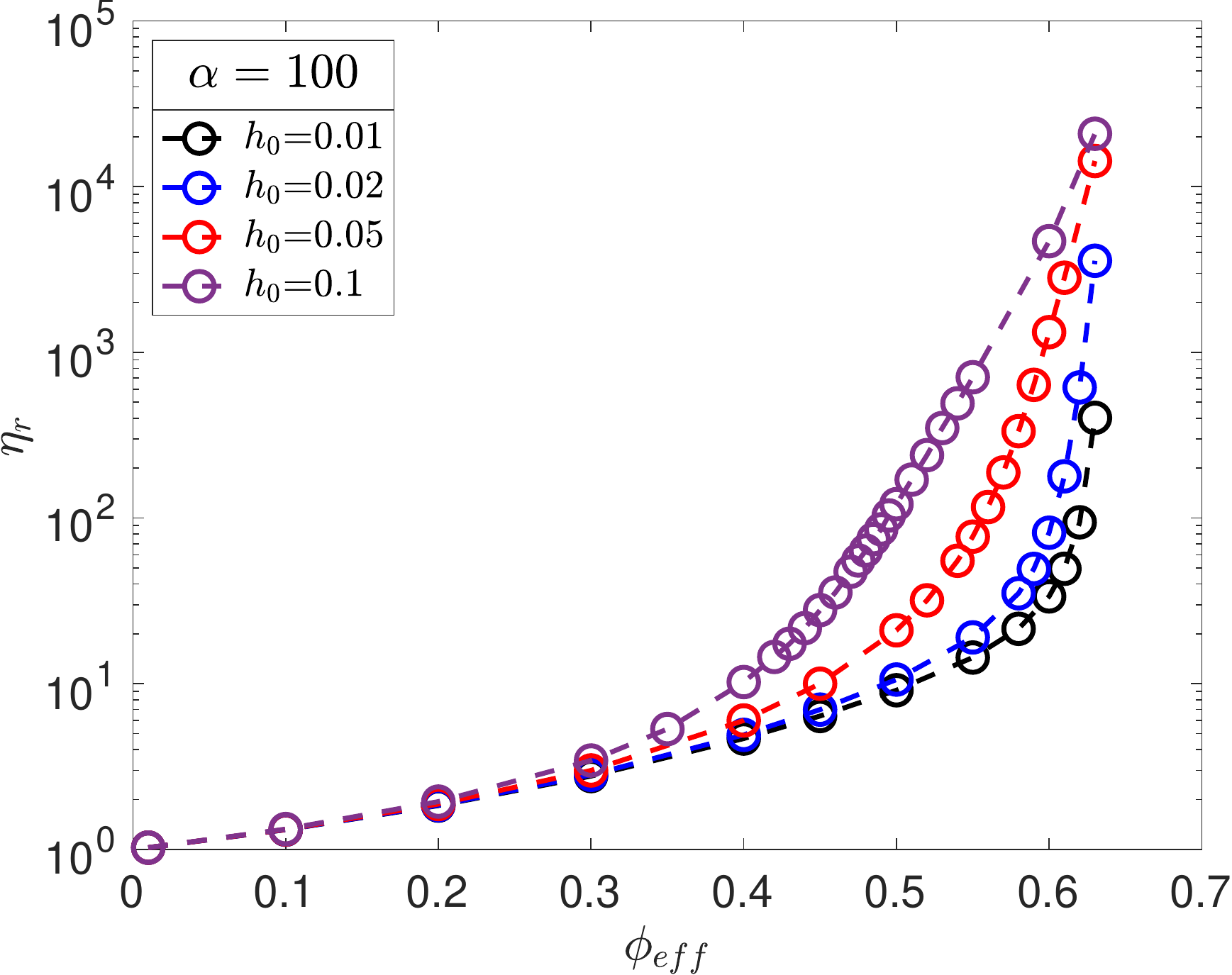}
         \caption{}
         \label{fig:etaralpha100}
     \end{subfigure}
     \begin{subfigure}[b]{0.4\textwidth}
         \centering
         \includegraphics[width=\textwidth]{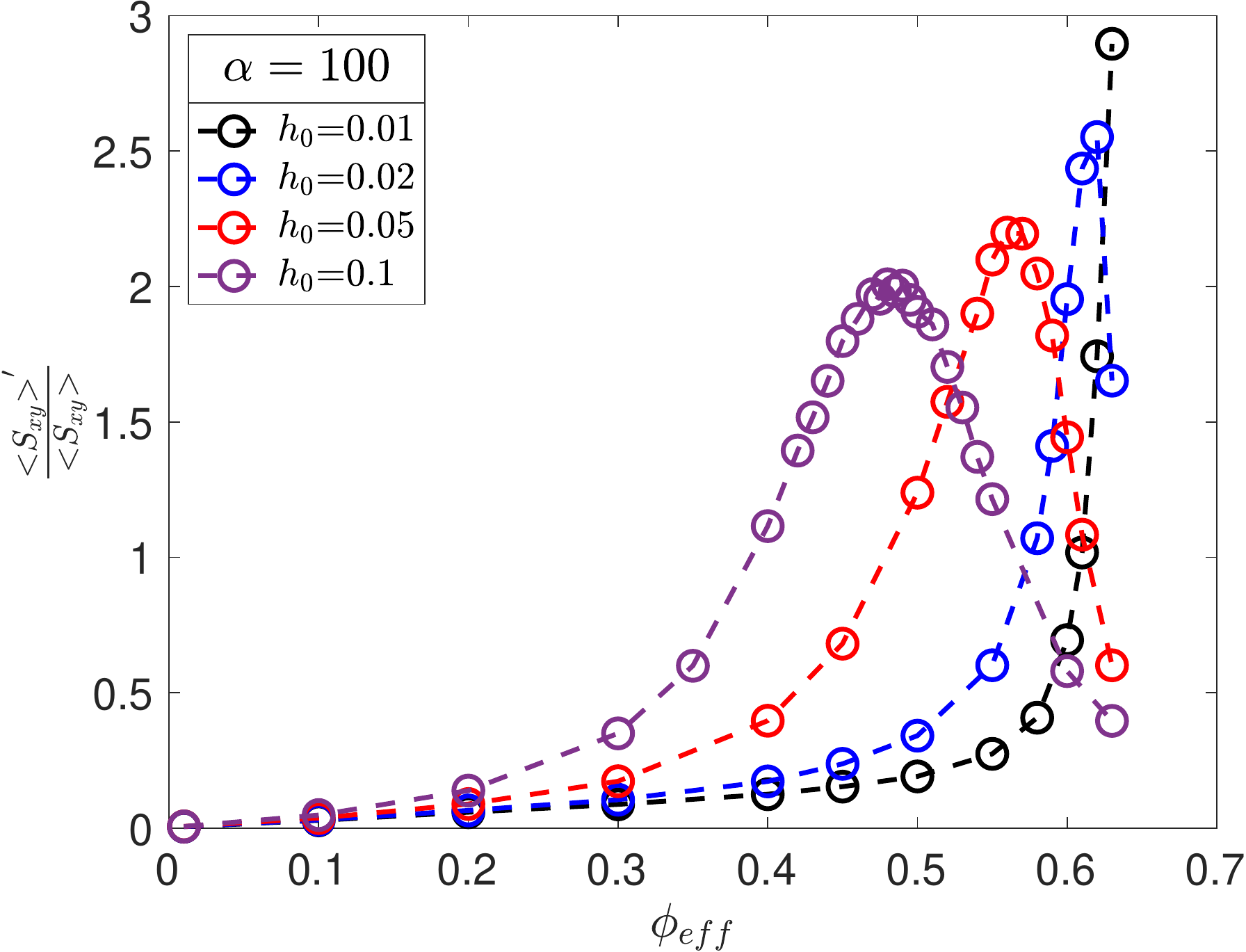}
         \caption{}
         \label{fig:Sxyfluctalpha100}
     \end{subfigure}
     \begin{subfigure}[b]{0.4\textwidth}
         \centering
         \includegraphics[width=\textwidth]{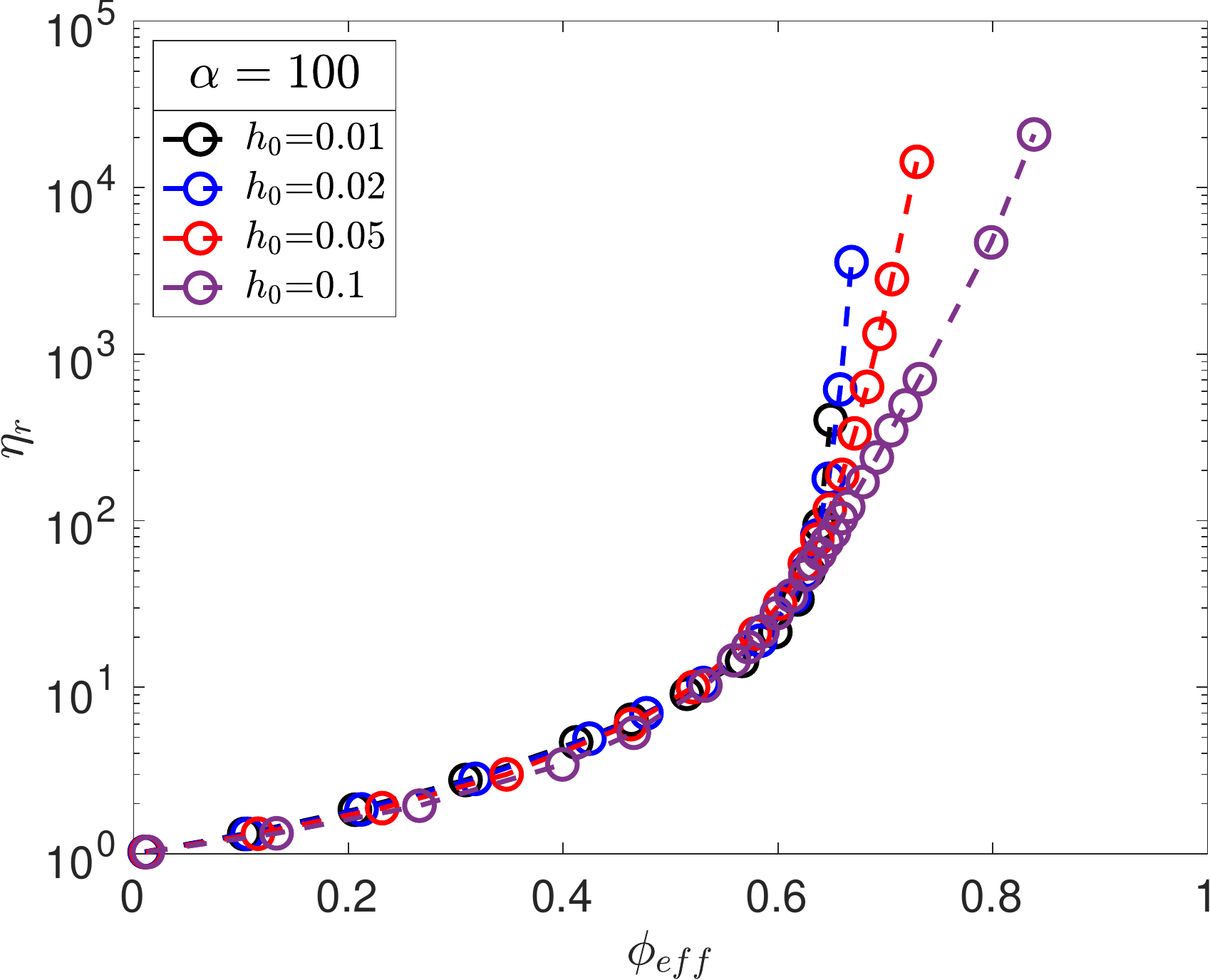}
         \caption{}
         \label{fig:etaralpha100norm}
     \end{subfigure}
     \begin{subfigure}[b]{0.4\textwidth}
         \centering
         \includegraphics[width=\textwidth]{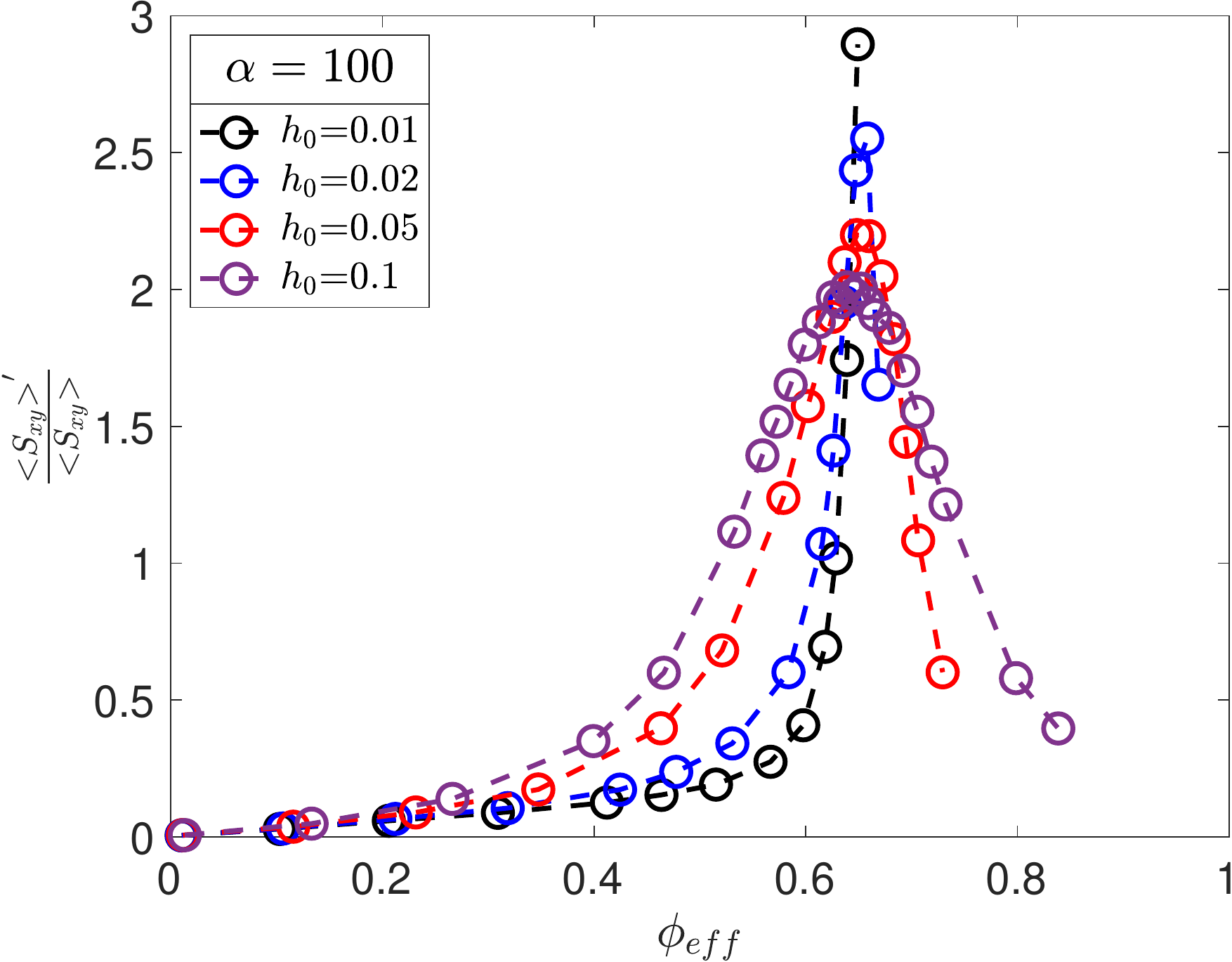}
         \caption{}
         \label{fig:Sxyfluctalpha100norm}
     \end{subfigure}
        \caption{($a$) and ($b$)  High frequency viscosity $\eta_r$ and  normalized fluctuations in stresslet $S_{xy}$ of a suspension as a function of particle volume fraction when friction range is varied between $h_0/a=0.01$ and $h_0/a=0.1$ and friction strength $\alpha=100$. The particle volume fraction is scaled with friction range $\phi_{eff}=\phi(1+h_0/a)^3$ in ($c$) and ($d$).}
        \label{fgr:etarSxyflucth0vary}
\end{figure}

The stress fluctuations reveal an interesting consequence of the friction model based on enhanced hydrodynamic resistance. At low friction range $h_0/a=0.01$, the normalized fluctuations in the stresslet increase weakly with concentration for most of the concentration range simulated and increases rapidly at higher concentrations near random closed packing. The rate of increase in the later region increases with increase in $\alpha$. At $\phi=0.63$, for example, the normalized fluctuations in stresslet for frictional case with $\alpha=100$ are one order of magnitude higher compared to the frictionless case. The trend qualitatively matches the behavior of corresponding $\eta_r$ versus $\phi$ curve. However, the behaviour is quite different for $h_0/a=0.1$ case. For all the cases of $\alpha$ considered here, the viscosity  monotonically increases with $\phi$ but the normalized stresslet fluctuations first increase with $\phi$ and reach a peak value at a particular concentration $\phi=\phi_p$ after which they decay rapidly to reach a value apparently independent of $\alpha$ at $\phi=0.63$. Note that the fluctuations of the frictional case here are still larger than that of the frictionless case. As the particle concentration increases beyond $\phi_p$ the number of particle pairs with surface separation below $h_0/a$ increases significantly and the hydrodynamic coupling between the particles increases due to the enhanced frictional resistance to sliding both in approach and departure. This coupling results in reduction of fluctuations in stresslet with increase in $\phi$ beyond $\phi_p$. Increase in $\alpha$ is observed to produce larger fluctuations at all concentrations and faster rate of increase of the fluctuations near $\phi=\phi_p$. The $\phi_p$ corresponding to peak fluctuations is observed to change weakly with $\alpha$: $\phi_p=0.481$, $0.487$ and $0.516$ for $\alpha=1$, $10$ and $100$ respectively. However it strongly depends on $h_0/a$. 

Figures \ref{fig:etaralpha100} and \ref{fig:Sxyfluctalpha100} shows the dependence of high frequency viscosity and normalized stresslet fluctuations on the friction range $h_0/a$ for $\alpha=100$. As $h_0/a$ is increased from $h_0/a=0.01$ to $h_0/a=0.1$, the magnitude of peak fluctuations reduced by two thirds, the peak widened and $\phi_p$ corresponding to the peak fluctuations reduced from $\phi_p\approx0.63$ to $\phi_p\approx0.49$. The viscosity at higher concentrations increased with increase in $h_0/a$ and, interestingly, the shape of the $\eta_r$ versus $\phi$ curve changes. At smaller friction ranges, $h_0/a=0.01$ for example, the $\eta_r$ follows a power-law dependence on $\phi$. Where as for higher values of $h_0/a$, for example at $h_0/a=0.1$, $\eta_r$ versus $\phi$ transitions from power-law behaviour for $\phi<\phi_p$ to exponential in nature for $\phi>\phi_p$ which, interestingly, is reminiscent of the percolation phase transition. The suspension phase transition here is driven by the frictional hydrodynamic resistance to sliding which is similar in nature to the lubrication resistance during squeezing.

When the friction range is incorporated into the particle size to account for the effect of size of asperities, the effective particle volume fraction increases, $\phi_{eff} = \phi(1+h_0/a)^3$. This scaling of the concentration, as shown in figures \ref{fig:etaralpha100norm} and \ref{fig:Sxyfluctalpha100norm}, aligns the peaks of the fluctuation curves for different $h_0/a$ at the same value of the effective concentration  $\phi_{effp}\approx65$ which is close to the random packing fraction of smooth hard spheres, $\phi_r$. The scaling also resulted in collapse of $\eta_r$ curves for different $h_0/a$ on to a master curve for $\phi_{eff}<\phi_{effp}$. Note that the error bars in figures \ref{fgr:etarSxyfluctalphavary} and \ref{fgr:etarSxyflucth0vary} are smaller than the marker size and are not shown.

The coarse grained hydrodynamic model developed here for rough suspensions requires friction strength, $\alpha$, and friction range relative to the hydrodynamic radius of the particle, $h_0/a$, as inputs which can be estimated experimentally for real-world particles as described in the previous section. The model captures the hydrodynamic contribution arising from the interaction of rough particle pairs with average surface separation less than the friction range. For rough particle suspensions with effective volume fraction less than the random closed packing, $\phi(1+h_0/a)^3<\phi_r$, the model predicts a high frequency viscosity that diverges with a power-law dependence on the particle volume fraction. Where as for $\phi(1+h_0/a)^3>\phi_r$, the model predicts a viscosity with exponential dependence on the particle loading.

\section{Conclusions}
A coarse grained hydrodynamic model is parameterized to simulate friction between rough particle pairs in suspension flows. The model requires an input of two free parameters: friction coupling range relative to the particle size, $h_0/a$, and friction coupling strength, $\alpha$. In the model, the resistance contributions from various modes of sliding and rolling motions of the particle pair were strengthened from the weakly diverging $O($log$(1/h))$ form to a strongly diverging $O(1/h)$ form where $h$ is the average particle surface separation. Relative magnitudes between various frictional resistance modes were derived to satisfy the positive semi-definiteness of the grand resistance matrix and to satisfy the rigid dumbbell-like motion of the rough particle pair at contact and at the limit of large friction strength. 

The hydrodynamic contribution to the high frequency viscosity of a suspension of rough particles as a function of particle roughness, $\alpha$ and $h_0/a$ is computed using Stokesian dynamics simulations integrated with the hydrodynamic friction model. In the vanishing roughness limit, the model recovers the smooth particle suspension viscosity which is in good agreement with the results of Stokesian dynamics simulations of Sierou and Brady~\cite{sierou2001accelerated}. With increase in particle roughness or $\alpha$ and $h_0/a$, the model predicts increase in viscosity which diverges with a power-law dependence on the particle volume fraction, $\phi$, for $\phi(1+h_0/a)^3<\phi_r$, where $\phi_r$ is the random closed packing for smooth hard spheres. This is a characteristic of hard sphere suspensions. Where as for $\phi(1+h_0/a)^3>\phi_r$, interestingly, the additional frictional hydrodynamic contributions of neighbouring particles increase the particle coupling. In this regime, the model predicts viscosity with exponential dependence on $\phi$ which is a characteristic of percolating suspensions. 

In addition, we proposed a method to estimate the free parameters of the hydrodynamic friction model for real-world rough suspensions. For that purpose, dynamics of a pair of rough particles subjected to a linear shear flow were simulated using Stokesian dynamics tool integrated with the hydrodynamic friction model. A simple model for the relative rotation rate of the rough pair was constructed from the simulated trajectories as a function of particle orientation, inter-particle separation and the particle roughness via the two parameters of the friction model. One can experimentally measure the rotation rates of particles in linear shear flow and compare them with the pair rotation rates computed from the model to estimate the values for the two parameters of the hydrodynamic friction model. We outlined the steps involved in such estimation.
 
 Although present work considered non-Brownian particles, the hydrodynamic model presented here can be used to simulate colloidal suspensions of rough particles. This is because the additional hydrodynamic resistance contributions due to particle roughness and the smooth particle lubrication resistance functions are of the same nature and the two contributions are linearly added before computing Brownian displacements in Stokesian dynamics simulations of rough colloids. This guarantees a thermodynamically consistent sampling of Brownian displacements for rough colloids.

\bibliographystyle{ieeetr}
\bibliography{rsc}

\end{document}